\documentclass[sn-mathphys-num]{sn-jnl}

\usepackage{multirow}%
\usepackage[nohyperlinks]{acronym}
\usepackage{algorithm}%
\usepackage{algorithmicx}%
\usepackage{algpseudocode}%
\usepackage{amsmath,amssymb,amsfonts}%
\usepackage{amsthm}%
\usepackage[title]{appendix}%
\usepackage{booktabs}%
\usepackage{graphicx, import}%
\usepackage[utf8]{inputenc}
\usepackage[T1]{fontenc}
\usepackage{listings}%
\usepackage{lmodern}
\usepackage{mathrsfs}%
\usepackage{manyfoot}%
\usepackage{natbib}
\usepackage[separate-uncertainty=true]{siunitx}
\usepackage{tabularx}
\usepackage{textcomp}%
\usepackage{xcolor}%
\usepackage[colorinlistoftodos, prependcaption, textsize = tiny]{todonotes}	
\usepackage{soul}
\usepackage{tikz}
\usetikzlibrary{arrows,arrows.meta}
\usepackage{epstopdf}

\acrodefplural{AOM}[AOMs]{acousto-optic modulators}
\acrodefplural{NTC}[NTCs]{negative temperature coefficient thermistors}
\acrodefplural{fAOM}[fAOMs]{fibre based acousto-optical modulators}
\acrodefplural{PCB}[PCBs]{printed circuit boards}
\acrodefplural{MS}[MSs]{mating sleeves}

\theoremstyle{thmstyleone}%
\theoremstyle{thmstyletwo}%

\theoremstyle{thmstylethree}%

\begin{document}

\title[BECCAL LS Design Paper]{
	 Comparison of laser system designs for quantum technologies: BECCAL flight system vs. BECCAL ground test bed}

\author*[1,2,a]{\fnm{Victoria A.} \sur{Henderson}}\email{vicki.henderson@stfc.ac.uk}
\author[3]{\fnm{Jean-Pierre} \sur{Marburger}}\email{jean-pierre.marburger@uni-mainz.de}
\author[3]{\fnm{André} \sur{Wenzlawski}}\email{awenzlaw@uni-mainz.de}
\author[1,2]{\fnm{Tim} \sur{Kroh}}\email{kroh@physik.hu-berlin.de}
\author[1]{\fnm{Hamish} \sur{Beck}}\email{hamish@physik.hu-berlin.de}
\author[1]{\fnm{Marc} \sur{Kitzmann}}\email{marc.kitzmann@physik.hu-berlin.de}
\author[2]{\fnm{Ahmad} \sur{Bawamia}}\email{ahmad.bawamia@fbh-berlin.de}
\author[4]{\fnm{Marvin} \sur{Warner}}\email{marvin.warner@zarm.uni-bremen.de}
\author[4]{\fnm{Mareen L.} \sur{Czech}}\email{mareen.czech@zarm.uni-bremen.de}
\author[1]{\fnm{Matthias} \sur{Schoch}}\email{mschoch@physik.hu-berlin.de}
\author[1,2]{\fnm{Jakob} \sur{Pohl}}\email{jakob.pohl@physik.hu-berlin.de}
\author[2]{\fnm{Matthias} \sur{Dammasch}}\email{matthias.dammasch@fbh-berlin.de}
\author[2]{\fnm{Christian} \sur{K\"{u}rbis}}\email{christian.kuerbis@fbh-berlin.de}
\author[5]{\fnm{Ortwin} \sur{Hellmig}}\email{ortwin.hellmig@uni-hamburg.de}
\author[1,b]{\fnm{Christoph} \sur{Grzeschik}}\email{grzeschik@iit-berlin.de}
\author[1,2]{\fnm{Evgeny V.} \sur{Kovalchuk}}\email{evgeny.kovalchuk@physik.hu-berlin.de}
\author[1]{\fnm{Bastian} \sur{Leykauf}}\email{leykauf@physik.hu-berlin.de}
\author[1]{\fnm{Hrudya} \sur{Thaivalappil Sunilkumar}}\email{hrudya.thaivalappil.sunilkumar@physik.hu-berlin.de}
\author[1]{\fnm{Christoph} \sur{Weise}}\email{christoph.weise@physik.hu-berlin.de}
\author[3]{\fnm{S\"{o}ren} \sur{Boles}}\email{soboles@uni-mainz.de}
\author[3]{\fnm{Esther} \sur{del Pino Rosendo}}\email{esdelpin@uni-mainz.de}
\author[3]{\fnm{Faruk A.} \sur{Sellami}}\email{fsellami@uni-mainz.de}
\author[5]{\fnm{Bojan} \sur{Hansen}}\email{asmus.bojan.hansen@uni-hamburg.de}
\author[2]{\fnm{Jan M.} \sur{Baumann}}\email{janmarkus.baumann@fbh-berlin.de}
\author[2]{\fnm{Tobias} \sur{Franke}}\email{tobias.franke@fbh-berlin.de}
\author[2]{\fnm{Alina} \sur{Hahn}}\email{alina.hahn@fbh-berlin.de}
\author[2]{\fnm{Karl} \sur{H\"{a}usler}}\email{karl.Haeusler@fbh-berlin.de}
\author[2]{\fnm{Max} \sur{Schiemangk}}\email{max.schiemangk@fbh-berlin.de}
\author[2]{\fnm{Robert} \sur{Smol}}\email{robert.smol@fbh-berlin.de}
\author[2]{\fnm{Jonas} \sur{Strobelt}}\email{jonas.strobelt@fbh-berlin.de}
\author[5]{\fnm{Klaus} \sur{Sengstock}}\email{klaus.sengstock@uni-hamburg.de}
\author[2]{\fnm{Andreas} \sur{Wicht}}\email{andreas.wicht@fbh-berlin.de}
\author[3]{\fnm{Patrick} \sur{Windpassinger}}\email{windpass@uni-mainz.de}
\author[1,2]{\fnm{Achim} \sur{Peters}}\email{achim.peters@physik.hu-berlin.de}

\affil*[1]{\orgdiv{Institut f\"{u}r Physik}, \orgname{Humboldt-Universit\"{a}t zu Berlin}, \orgaddress{\street{Newtonstr. 15}, \city{Berlin}, \postcode{12489}, \state{Berlin}, \country{Germany}}}
\affil[2]{\orgname{Ferdinand-Braun-Institut, Leibniz-Institut für Höchstfrequenztechnik}, \orgaddress{\street{Gustav-Kirchoff-Str. 4}, \city{Berlin}, \postcode{12489}, \state{Berlin}, \country{Germany}}}
\affil[3]{\orgdiv{Institut f\"{u}r Physik}, \orgname{Johannes Gutenberg University Mainz}, \orgaddress{\street{Staudingerweg 7}, \city{Mainz}, \postcode{55128}, \state{Rhineland-Palatinate}, \country{Germany}}}
\affil[4]{\orgdiv{ZARM, Universit\"{a}t Bremen}, \orgaddress{\street{Am Fallturm 2}, \city{Bremen}, \postcode{28359}, \state{Bremen}, \country{Germany}}}
\affil[5]{\orgdiv{Institut f\"{u}r Quantenphysik}, \orgname{Universität Hamburg}, \orgaddress{\street{Luruper Chaussee 149}, \city{Hamburg}, \postcode{22761}, \state{Hamburg}, \country{Germany}}}
\affil[a]{\orgname{Now at RAL Space, Science Technology and Facilities Council}, \orgaddress{\street{Rutherford Appleton Laboratory}, \city{Harwell, Didcot},  \state{Oxfordshire}, \postcode{OX11 0QX}, \country{UK}}}
\affil[b]{\orgname{Now at Institute for Innovation and Technology}, \orgaddress{\street{Steinplatz 1}, \city{Berlin}, \postcode{10623}, \state{Berlin}, \country{Germany}}}

\abstract{We present the design of laser systems for the \ac{BECCAL} payload, enabling numerous quantum technological experiments onboard the \ac{ISS}, in particular dual species $^{87}$Rb and $^{41}$K Bose-Einstein condensates.  A \ac{FM} and a \ac{COTS} based model are shown, both of which meet the \ac{BECCAL} requirements in terms of functionality, but have differing \ac{SWaP} and environmental requirements.  The capabilities of both models are discussed and characteristics compared.

The flight model of \ac{BECCAL} uses specifically developed and qualified custom components to create a compact and robust system suitable for long-term remote operation onboard the \ac{ISS}.  This system is based on \acs{ECDL}-\acs{MOPA} lasers and free-space optical benches made of Zerodur, as well as commercial fibre components. The \ac{COTS}-based system utilizes entirely commercial parts to create a functionally equivalent system for operation in a standard laboratory, without the strict \ac{SWaP} and environmental constraints of the flight model.}

\keywords{Laser system, Quantum technology, Microgravity, International Space Station, Bose-Einstein Condensate}


\maketitle

\section{Introduction}\label{sec:intro}

As atomic physics technology matures and more ambitious measurements are considered, we are moving towards an advent of facility-scale and multi-user cold atom apparatuses both on ground and in space~\cite{Bongs2019,Alonso2022,Abend2023}. This includes the \ac{ISS} payload considered in this paper: \ac{BECCAL}~\cite{Frye2021}.

Such facilities and experiments target a variety of science objectives including Einstein equivalence principle tests and other tests of fundamental physics, the observation of gravitational waves and the detection of ultra-light dark matter.
For example, MIGA~\cite{Sabulsky2020, Canuel2022}, AION~\cite{Badurina2020}, ZAIGA~\cite{Zhan2020} and ELGAR~\cite{Canuel2020} are all proposed or in production on-ground atom interferometry systems which target the infrasound band for gravitational waves among other capabilities.
Operation in micro-gravity enables the exploitation of different techniques and measurement regimes compared to those possible in gravity, with the ideal situation being perpetual space where one can access unlimited micro-gravity time with a stepping stone provided by platforms such as sounding rockets, Einstein elevators or drop towers.
Orbital proposals include AEDGE~\cite{El-Neaj2020}, CARIOQA~\cite{Leveque2022} and STE-QUEST~\cite{Ahlers2022} and far future capabilities in the future quantum explorer~\cite{Thompson2022}.

Robust, reliable, compact, reproducible, and resource efficient laser systems are a crucial part of current developments.
Often multiple duplicate laser systems are required, and the intended environment can result in strong requirements concerning \ac{SWaP}, vibration or shock immunity, resilience under thermal loads and cycling, and the ability to operate for long periods in remote or inaccessible places.
Extensive progress has been made in the facilities already in progress~\cite{Sabulsky2020}, and in planned and operating orbital missions such as \ac{CAL}~\cite{Aveline2020}, CACES~\cite{Liu2018}, ACES~\cite{Laurent2015}, CAPR~\cite{Li2023}, and \ac{BECCAL}~\cite{Frye2021}.
This ongoing progress is supported and complemented by active and heritage experiments on micro-gravity platforms such as: the sounding rocket missions MAIUS~\cite{Schkolnik2016, Becker2018a}, JOKARUS~\cite{Doringshoff2019,Schkolnik2017}, FOKUS~\cite{Lezius2016}, and KALEXUS~\cite{Dinkelaker2017}; drop tower experiments such as QUANTUS~\cite{Rudolph2015, Deppner2021} and PRIMUS~\cite{Vogt2020}; parabolic flights~\cite{Barrett2016}; and recently commissioned Einstein Elevators~\cite{Condon2019, Raudonis2023, Pelluet2024}

\ac{BECCAL} is planned as a multi-user and multi-purpose experiment onboard the \ac{ISS} targeting a multitude of scientific questions in different areas such as atom interferometry, spinor condensates, and quantum mixtures, utilising rubidium and potassium. Of particular relevance to laser system design, it will offer higher atom numbers, faster cycle rates and more complex optical traps as well as improved atom-interferometry abilities compared to prior payloads. 

The payload consists of three lockers to be inserted into an \ac{ISS} \ac{EXPRESS} rack: a single locker containing control electronics; a dual locker containing the physics package and associated electronics; and a dual locker containing the laser system and associated electronics. 
The conceptual details of the payload have already been described in previous work~\cite{Frye2021}.
\ac{BECCAL} builds directly on the heritage of the drop tower and sounding rocket missions MAIUS and QUANTUS~\cite{Schkolnik2016, Becker2018a, Kulas2017}.

In this paper we present the laser system designs of the \ac{BECCAL} payload.  
This includes two designs which fulfil the \ac{BECCAL} laser system science requirements: one \ac{FM} system which uses specifically developed and qualified custom components to create a compact and robust system suitable for the \ac{ISS}; and one which uses \ac{COTS} components to create a system suitable for use in a standard lab. 
In Sec.~\ref{sec:Req}, the specific requirements of the \ac{BECCAL} laser system are described, expanding on the information provided in~\cite{Frye2021}.
The ISS-suitable flight model design will be presented in Sec.~\ref{sec:FM}.
The \ac{COTS}-based laboratory model will be presented in Sec.~\ref{sec:GTB}, and finally, in Sec.~\ref{sec:Comp}, the two models will be compared.
We note that the laser systems are considered in isolation without the associated control electronics.

\section{An overview of laser system requirements} \label{sec:Req}

The \ac{BECCAL} laser system has to comply to a wide range of requirements which are necessary to facilitate the comprehensive functionality of the \ac{BECCAL} payload, which is described in detail in~\cite{Frye2021}. In short, in \ac{BECCAL} it is possible to produce individual and dual species $^{87}$Rb and $^{41}$K BECs as well as ultra-cold $^{85}$Rb, $^{39}$K, and $^{40}$K, to perform Raman interferometry on two axes, and to trap in a red- or blue-detuned dipole trap.
These key system requirements result in comprehensive requirements on the laser system sub-system, including laser light at \SI{780}{\nano\meter} and \SI{767}{\nano\meter} for cooling and control of rubidium and potassium, as well as \SI{1064}{\nano\meter} and \SI{764}{\nano\meter} for dipole trapping. This light is provided to the physics package in \num{15} optical fibres bridging the two dual-locker systems. 
A summary of the key functionalities and capabilities of the \ac{BECCAL} laser system is given in Tab.~\ref{tab:OverviewTable}.

\begin{table}[ht!]
	\caption{Overview of the experimental capabilities of the \ac{BECCAL} laser system.}
	\label{tab:OverviewTable}
	\begin{tabular}{p{0.5\textwidth}p{0.45\textwidth}}
		\hline	
		& \\
		\textbf{Total laser powers} & \\
		Rb 3D-MOT cooling (repump) - 4 beams & $\geq$ \SI{90}{\milli\watt} ($\geq$ \SI{12}{\milli\watt}) \\
		Rb 2D-MOT cooling (repump) - 4 beams & $\geq$ \SI{80}{\milli\watt} ($\geq$ \SI{40}{\milli\watt}) \\
		Rb Detection cooling (repump) - 2 beams & $\geq$ \SI{6}{\milli\watt} ($\geq$ \SI{2}{\milli\watt}) \\
		K 3D-MOT cooling (repump) - 4 beams & $\geq$ \SI{75}{\milli\watt} ($\geq$ \SI{65}{\milli\watt}) \\
		K 2D-MOT cooling (repump) - 4 beams & $\geq$ \SI{70}{\milli\watt} ($\geq$ \SI{70}{\milli\watt}) \\
		K Detection cooling (repump) - 2 beams & $\geq$ \SI{7}{\milli\watt} ($\geq$ \SI{7}{\milli\watt}) \\
		Rb Interferometry primary (secondary) & $\geq$ \SI{10}{\milli\watt} ($\geq$ \SI{10}{\milli\watt}) \\
		K Interferometry primary (secondary) & $\geq$ \SI{10}{\milli\watt} ($\geq$ \SI{10}{\milli\watt}) \\
		1064nm Dipole - 2 beams & $\geq$ \SI{300}{\milli\watt} \\
		764nm Dipole - 2 beams & $\geq$ \SI{40}{\milli\watt} \\
		& \\
		\textbf{Frequency control} & \\
		Frequency adjustment rate & $\geq$ \SI{1}{\giga\hertz\per\milli\second} \\
		Frequency resolution & $\leq$ \SI{100}{\kilo\hertz} \\	
		Relative frequency uncertainty for Rb and K light & $\leq$ \num{3e-9} \\
		FWHM linewidth for Rb, K, and \SI{1064}{\nano\meter} light & $\leq$ \SI{100}{\kilo\hertz} (\SI{1}{\milli\second}) \\		
		Linewidth of \SI{764}{\nano\meter} light & $\leq$ \SI{10}{\mega\hertz} ($\leq$ \SI{3e6}{\hertz\squared\per\hertz} for frequencies above~\SI{100}{\milli\hertz}) \\
		Wavelength of \SI{1064}{\nano\meter} light & \SI{1064\pm10}{\nano\meter} \\		
		& \\
		\textbf{Intensity control} & \\
		Extinguishing via \acs{AOM}  & By \SI{-30}{\decibel} in $\leq$ \SI{10}{\micro\second} \\
		Extinguishing via shutter  & $\leq$ \SI{10}{\milli\second} \\
		Linear ramp & \SIrange{0.1}{100}{\percent} in \SI{1}{\milli \second} \\
		Stabilization before first switching element &  $\leq$ \SI{0.1}{\percent} with a bandwidth $\geq$ \SI{100}{\hertz} \\
		Measurement of power in physics package & Before each experimental run with a resolution of $\leq$ \SI{0.1}{\percent} \\
		& \\
		\hline
	\end{tabular}
\end{table}

There will be at least three \ac{BECCAL} models: two ground testbeds (one each serving the US and Germany), and a flight model.
Additionally, outside of the \ac{BECCAL} collaboration, scientists may also wish to reproduce the \ac{BECCAL} laser system functionality for various purposes.
One of the ground testbeds will consist of \ac{COTS} components in such a way as to replicate the \ac{BECCAL} flight model as closely as possible without the same restrictive \ac{SWaP} constraints of an ISS \ac{EXPRESS} locker.
As detailed previously, both the \ac{COTS} model and \ac{FM} will conform to the same system and sub-system requirements, and will interface to the other sub-systems (such as electronics) as similarly as possible.

Several approaches can be taken to produce the light fields required for experiments such as \ac{BECCAL}~\cite{Abend2023}.
One could exploit the qualification level of telecommunications C-band lasers and components by frequency doubling light at \SI{1560}{\nano\meter}, an approach taken in~\cite{Sabulsky2020} combined with free-space optical benches and in~\cite{Li2023} combined with fibre splitters.
Alternatively, as taken here and in the QUANTUS and MAIUS missions, micro-integrated diode lasers with integrated amplifiers can be used.
Due to the \ac{SWaP} budget of \ac{BECCAL}, optical power efficiency must be maximized in order to minimize the number of lasers used; this combined with the need to combine and split multiple wavelengths of light precisely, necessitates a combined free-space and fibred optical distribution system.
In the future, all-fibre distribution systems and photonic chips~\cite{Isichenko2022} may offer a further reduction to \ac{SWaP} budgets whilst still meeting scientific requirements.

\section{Design of the flight model laser system} \label{sec:FM}

\begin{figure}[ht!]
	\centering
	\includegraphics[width=\textwidth]{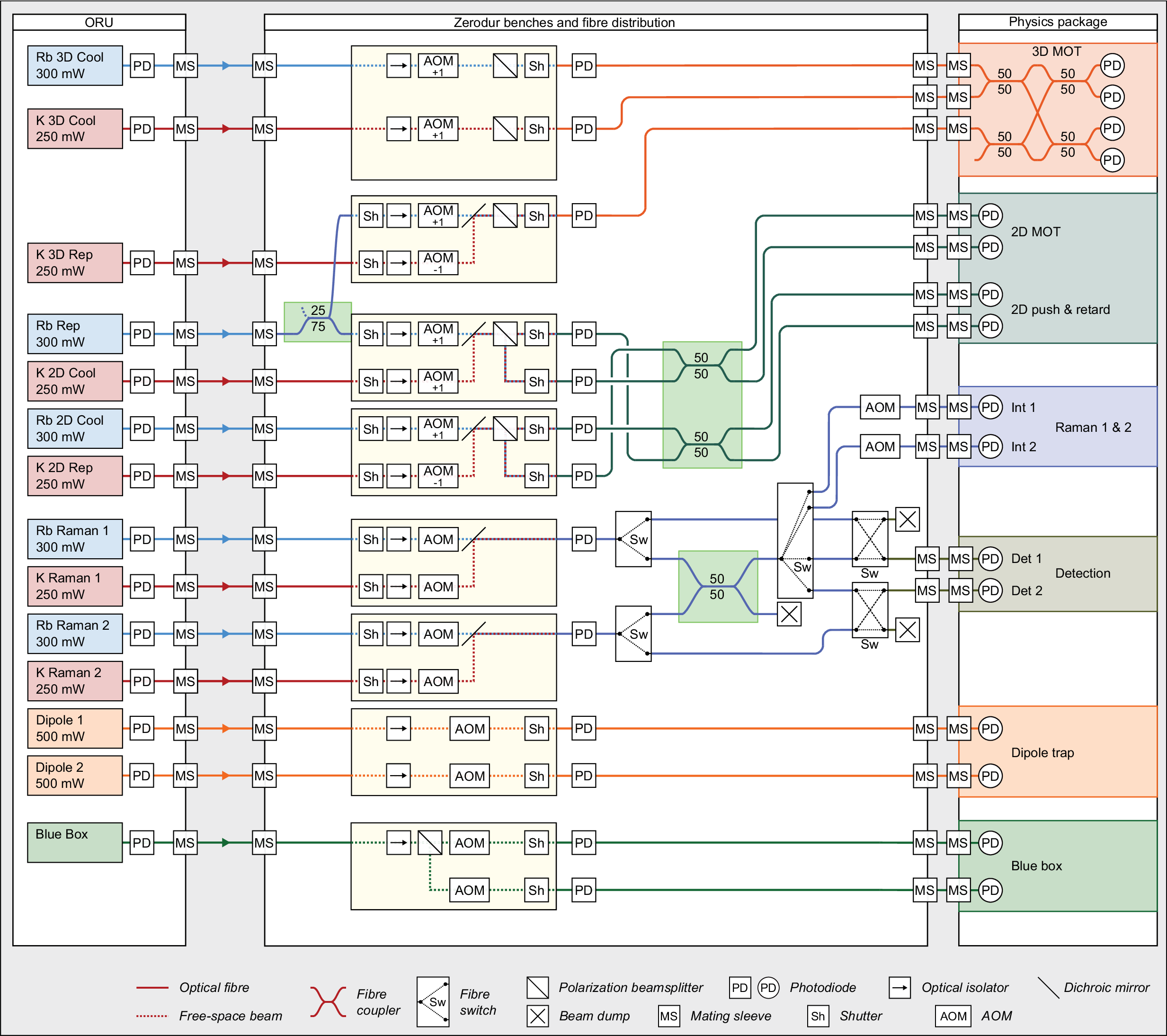}
	\caption{Optical schematic of the laser system.  Science lasers are shown on the left hand side, connected via inline \acs{PD}s and mating sleeves to the rest of the system.  Within the central panel, light is distributed via fibre optics and free-space optical benches (shaded yellow).  The right panel shows the delivery of light to the physics package via mating sleeves in function groups such as 3D-\ac{MOT} fibres.  Prior to the overlapping of multiple wavelengths, light paths are colour coded according to wavelength (blue, red, orange and green corresponding to \SI{780}{\nano\meter}, \SI{767}{\nano\meter}, \SI{1064}{\nano\meter}, and \SI{764}{\nano\meter} respectively).  After the free-space optical benches, the light paths are colour coded according to the functional groups delivered to the physics package.}
	\label{fig:overview}
\end{figure}

\begin{figure}[ht!]
	\centering
	\input{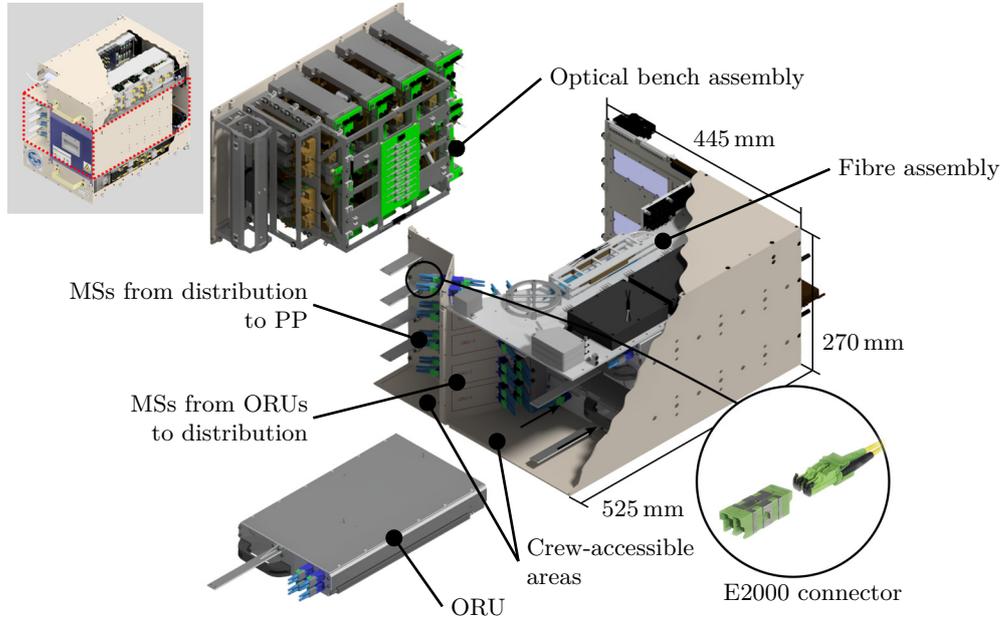}
	\caption{Renderings showing an overview of the \ac{BECCAL} flight laser system.  The inset in the top left corner shows the location of the laser system within the rest of the laser system double locker. Within the rendering, items (such as the optical bench assembly) are moved away from their actual position for visibility.}
	\label{fig:lockeroverview}
\end{figure}

The \ac{FM} laser system can be conceptually understood in three parts:  the laser modules in \acp{ORU}, free-space optics on Zerodur benches, and further distribution in fibre optics. These parts will be individually described in this section of the paper.
A schematic of the laser system is shown in Fig.~\ref{fig:overview} and a rendering is shown in Fig.~\ref{fig:lockeroverview}. 

The laser system contains 16 laser in total: 6 Rb lasers at \SI{780}{\nano\meter}; 7 K lasers at \SI{767}{\nano\meter}; 2 red-detuned dipole lasers at \SI{1064}{\nano\meter}; 1 blue-detuned dipole laser at \SI{764}{\nano\meter}.
One each of the Rb and K lasers are used as reference lasers, to which all other Rb and K lasers are frequency stabilized. The light of the other 14 lasers is guided to the physics package where it is used to cool and/or manipulate the atomic ensembles.
No frequency stabilization of the dipole lasers is required.

On the left hand side of the schematic, 14 \ac{BECCAL} \acs{ECDL}-\acs{MOPA} lasers (excluding reference lasers) are shown alongside inline \acp{PD} providing fast-feedback to stabilize the optical power.
These laser modules are housed in four sets of four in \acp{ORU}, which can be independently removed from the sub-system.

The \acp{ORU} are connected to the rest of the sub-system via mating sleeves and optical fibres in the crew accessible area of the locker.
To ensure crew safety, the crew accessible fibres are armoured, the front panel is light tight, and the area is protected by a removable cover which is only removed when the payload is unpowered.
Further details of the lasers and \acp{ORU} can be found in Sec.~\ref{sec:lasers}.

After the mating sleeves at the interface to the rest of the laser system, the light is guided to eight Zerodur distribution benches, also via optical fibres. On the benches the light is guided in free-space. The benches enable us to overlap, switch and split the different light fields needed for different purposes and atomic species in an efficient manner. They are detailed further in Sec.~\ref{sec:ZD}.

The light is then coupled back into optical fibres on the Zerodur benches, and another PD provides housekeeping data before the light is further split, combined, and switched in fibre optic components (see Sec.~\ref{sec:fibres}).

The light is then directed to the physics package via optical fibres and 15 mating sleeves.
As with the \ac{ORU} to laser system connection, the fibres are armoured, front panels are light tight, and the area is protected by a removable plate.

\subsection{Lasers in orbital replacable units} \label{sec:lasers}

\begin{figure}[ht!]
	\begin{tikzpicture}[>=stealth]
		\tikzset{every node}=[font=\normalsize]
		\node[inner sep=0pt] (Picture) at (0,0) {\includegraphics[height=0.29\textwidth]{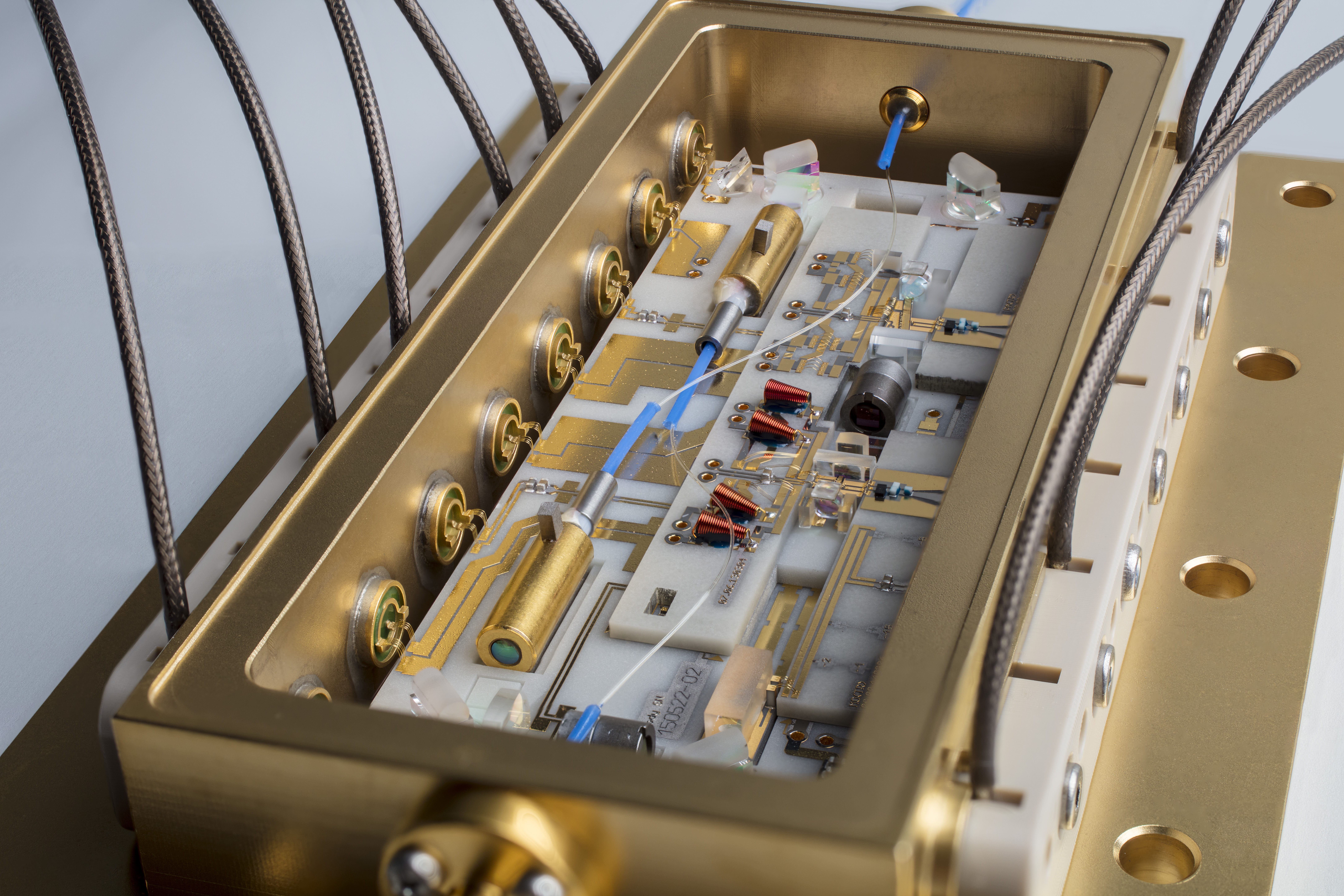}};
		\node[anchor=north west] at (-2.9,1.9) {\textbf{\textcolor{white}{(a)}}};
	\end{tikzpicture}
	\begin{tikzpicture}[>=stealth]
		\tikzset{every node}=[font=\small]
		\node[inner sep=0pt] (Picture) at (0,0) {\includegraphics[height=0.29\textwidth, trim=152 59 188 64, clip]{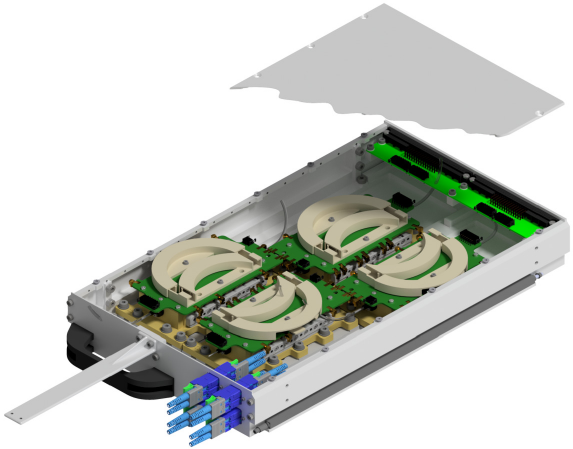}};
		
		\def\rend{5.4}
		\def\expr{-3.6}
		\def\lend{-2.5}

		\draw[{Circle[scale=1]}-,line width=1pt](.8,.7)--(-0.7,.9)node[left]{Interface PCB};		
		\draw[{Circle[]}-,line width=1pt](-1.3,-0.5)--(\lend,-0.195)node[left=0.2em,align=right]{Laser\\PCB};
		\draw[{Circle[]}-,line width=1pt](-1,-0.15)--(-1.7,0.4)node[left]{Fibre reel};
		\draw[{Circle[]}-,line width=1pt](-1.8,-1)--(\lend,-0.9)node[left]{Handle};
		\draw[{Circle[]}-,line width=1pt](-2.2,-1.55)--(\lend,-1.55)node[left=0.2em,align=right]{Storage\\rod};

		\draw[{Circle[]}-,line width=1pt](0.5,-0.9)--(1,-1.3)node[right,align=left]{\acs{ECDL}-\acp{MOPA}};
		\draw[{Circle[]}-,line width=1pt](1.3,-0.8)--(2.2,-0.7)node[right,align=left]{Wedge\\lock};
		\draw[{Circle[]}-,line width=1pt](-0.5,-1.6)--(0.5,-1.7)node[right]{Mating sleeves};
		\node[anchor=north west, font=\normalsize] at (-4.0,1.9) {\textbf{(b)}};
	\end{tikzpicture}
	\caption{{a) Photo of an \acs{ECDL}-\acs{MOPA} module similar to that used in \ac{BECCAL}~\cite{Kurbis2020}, the only difference is the differently shaped footprint. b) Rendering of an \ac{ORU} which houses four \acs{ECDL}-\acp{MOPA} in a removable drawer alongside breakout \acs{PCB}s and fibre optics.}}
	\label{fig:LaserModule}
\end{figure}
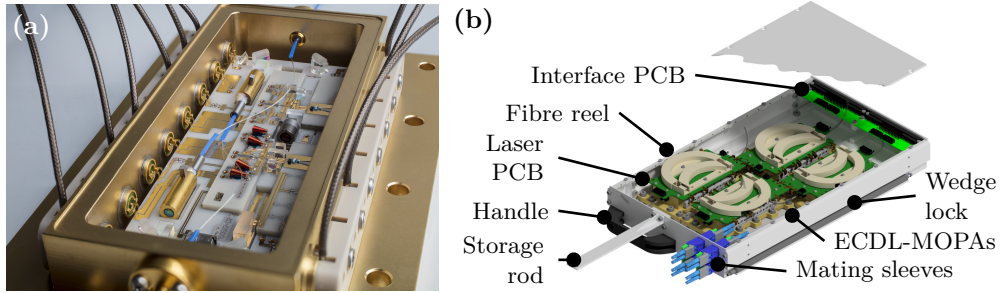

The laser sources used in the \ac{BECCAL} \ac{FM} are based on semiconductor laser chips in an \ac{ECDL} \ac{MOPA} configuration, as described in~\cite{Frye2021}. 
They are built into a fully-packaged laser module within a volume of \SI{125}{\milli\metre} x \SI{75}{\milli\metre} $\times$ \SI{22.5}{\milli\metre} and a total mass lower than \SI{0.8}{\kilogram}.
The power consumption of any single laser module within the operating conditions for \ac{BECCAL} is lower than \SI{5}{\watt}. 
Electrical signals are fed in and out of the module via Mini-SMP plugs, and two optical connections provide light via single-mode, polarisation-maintaining fibres. 
One of these optical fibres couples out the main optical path, which is then directed to the distribution part of the system. 
The second fibre is an auxiliary port which couples out the small amount of light at the back side of the ECDL chip.  This auxiliary light is generally used for laser locking within \ac{BECCAL}.
A detailed description of the laser module technology can be found in~\cite{Kurbis2020}. 
One such laser module is shown in Fig.~\ref{fig:LaserModule}(a), this photo shows a functionally identical module to those used in \ac{BECCAL}, however, in \ac{BECCAL} the footprint and mass is reduced via cut-outs.

The \ac{ECDL}-\ac{MOPA} laser modules are housed in groups of four within \acp{ORU}.
Should a problem occur with a single laser module, the use of \acp{ORU} enables us to replace a small sub-set of lasers, rather than replacing an entire locker.
The lasers are organized by wavelength such that maximum interchangeability is ensured.
Three \acp{ORU} contain two Rb and two K lasers each, and the fourth \ac{ORU} contains the remaining lasers, one K, two \SI{1064}{\nano\meter} and one \SI{764}{\nano\meter}.
A rendering of an \ac{ORU} is shown in Fig.~\ref{fig:LaserModule}(b). 
Here one can see the laser modules, associated breakout \acp{PCB}, and fibre management hardware. 

Electrical signals, such as currents for the laser chips, and signals for temperature stabilization and housekeeping, are routed in and out of the \acp{ORU} via connectors at the rear.
This means the \acp{ORU} are self-contained units which can be easily removed and replaced.
Passive housekeeping data includes temperatures and optical powers prior to fibre coupling inside the laser modules, thus enabling a level of diagnostics and health checks of the system.
The \ac{PCB} also houses an inline \ac{PD}, which is used to actively monitor and stabilize the laser output power within experimental cycles.
Further details of the intensity stabilization concept can be found in Sec.~\ref{sec:stab}.

Optical fibres are coiled within the \acp{ORU} before being routed through E2000 mating sleeves as previously described.
Outside the \acp{ORU}, crew-accessible optical fibres are coiled and secured to removable rods.

The \acp{ORU} are mechanically slotted into the locker structure and locked into place via wedge locks.
The bases of the \acp{ORU} make thermal contact with two water cooled heat sinks which are dedicated to their thermal management, with two \acp{ORU} being in contact with each of these heat-sinks.

\subsection{Free-space distribution on Zerodur benches} \label{sec:ZD}

For free-space light field distribution, \ac{BECCAL} makes use of fibre-coupled optical benches. 
Miniaturized optical components are glued onto baseplates made from Zerodur. 
These include custom designed and commercial optical components such as collimators and couplers as a fibre interface, \acp{AOM} and mechanical shutters for intensity switching, and dichroic mirrors and polarizing beam splitters to overlap light fields.

The BECCAL flight laser system features a total of eight benches for light field distribution, four of which have a footprint of $\SI{125}{\milli\meter}\! \times\! \SI{100}{\milli\meter}$ and the remaining four have a footprint of $\SI{125}{\milli\meter}\! \times\! \SI{120}{\milli\meter}$. All benches have a thickness of $\SI{30}{\milli\meter}$.

\paragraph{Bench toolkit}
To comply with the varying temperature conditions present at the ISS, the baseplate and many components are made from the glass ceramic Zerodur DK0S (produced by Schott AG), as this material features a coefficient of thermal expansion of only \SI{0\pm0.010 e-6}{\kelvin^{-1}}. 
Furthermore, it exhibits a density and elastic modulus comparable to aluminium, which ensures good mechanical stability.
In order to mount the miniaturized optics components to the baseplate, we use an adhesive bonding technique \cite{Duncker2014}. 
Optical benches made from the same toolkit were already employed in the sounding rocket missions FOKUS~\cite{Lezius2016}, KALEXUS~\cite{Dinkelaker2017}, MAIUS-1~\cite{Becker2018a,schkolnik_compact_2016} and the upcoming MAIUS-2/3~\cite{mihm_zerodur_2019,mihm_phd}.

The weight and size constraints for \ac{BECCAL} are even more stringent than for previous missions.
We have thus, in contrast to those missions, opted to place optical components on both sides of the optical bench. 

\begin{figure}[ht!]
	\centering
	\input{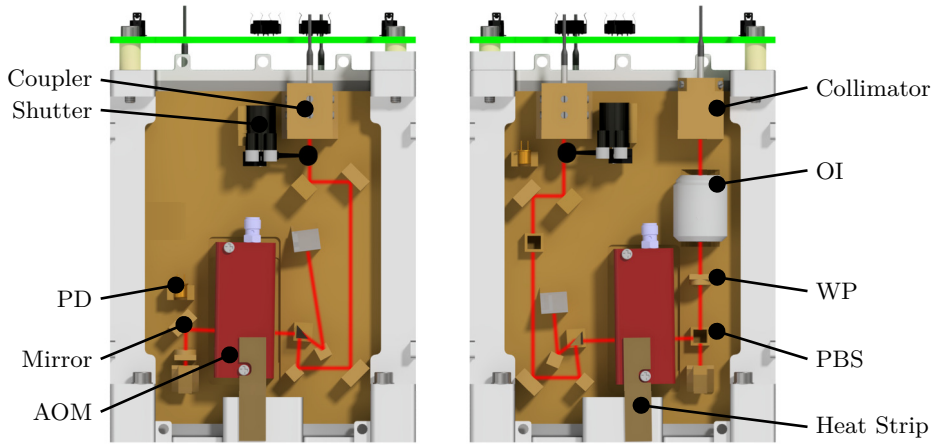}
	\caption{A top down rendering of both sides of an optical bench. On this particular bench, one light beam at \SI{780}{\nano\metre} (depicted in red) is split into two and then the intensities of both beams are independently controlled via two \acp{AOM} and shutters before the light is then coupled into two individual fibres.}\label{fig:bench_2d}
\end{figure}

\paragraph{Bench layout}
The layout of one of the optical benches is shown in Fig.~\ref{fig:bench_2d}.
On this bench, light enters through a fibre collimator, where light is collimated using a lens of focal length $f = \SI{4.51}{\milli\metre}$, which results in a collimated beam width of \SI{0.42\pm0.08}{\milli\metre} at \SIrange{764}{780}{\nano\metre} and \SI{0.58\pm0.11}{\milli\metre} at \SI{1064}{\nano\metre}. 
To suppress back reflections into the fibre, we use fibres with end-caps and \ac{AR} coatings, in addition to placing an \ac{OI} directly behind the collimator.

Free-space components are used to perform intensity control, as well as the merging and splitting of light fields. 
For fast intensity control, we use an \ac{AOM} in conjunction with a slower mechanical shutter for complete suppression of the light fields. 
To keep thermal loads to a minimum, we employ rotational solenoids instead of stepper motors to actuate the shutter blades, as they do not have to be continuously powered to retain their position.
Light fields of different frequencies are overlapped using narrow-band dichroic mirrors. 
This technique is used to overlap the light fields for cooling, trapping and interferometry at \SI{767}{\nano\metre} and \SI{780}{\nano\metre}, as can be seen in Fig.~\ref{fig:overview}. 
One beam can be split into multiple beams using \ac{PBS}. 
The splitting ratio can be adjusted using a \ac{WP}.

To guide the light between these components, we use mirrors made from Zerodur. 
Likewise, light is guided from one side of the optical bench to the other, using the same mirrors angled at \SI{45}{\deg}.

Light is coupled back into an optical fibre using an optical coupler, which contains the same lens as in the collimator. 
We regularly achieve fibre coupling efficiencies of above \SI{85}{\percent}~\cite{mihm_zerodur_2019}.

This approach to combine different functionalities in one optical bench is very efficient compared to using a series of fibred components to achieve the same result. 
This is because there is only one free-space to fibre transition, and so results in lower losses.

\begin{figure}[ht!]
	\centering
	\input{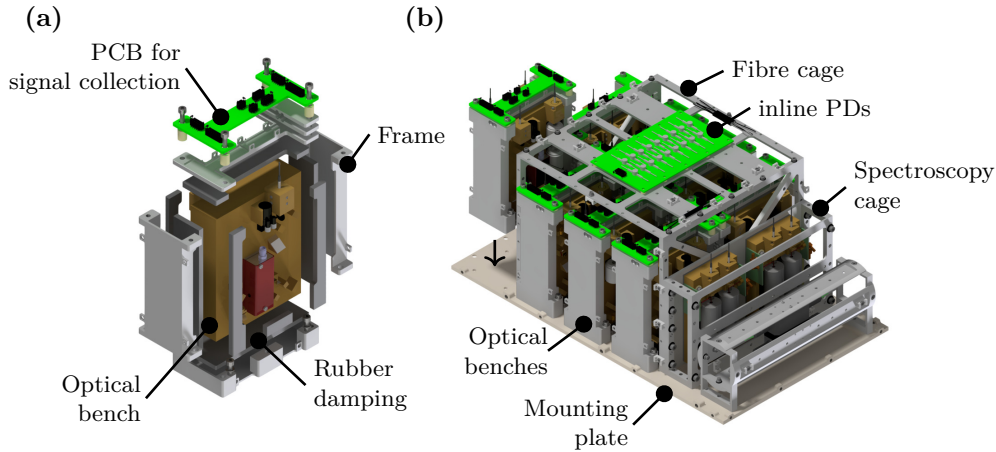}
	\caption{{a) Rendering of one Zerodur bench including the mounting frame, a rubber damping to minimize vibrations and mechanical stresson the Zerodur benches, and a PCB for collecting housekeeping signals. b) Rendering showing the integration of the Zerodur benches into the \ac{BECCAL} laser system. The Zerodur benches (including mounting frames) are screwed to a mounting plate where they are combined with the spectroscopy benches and inline photodiodes which are used for monitoring purposes.}}
	\label{fig:ZerodurBenches}
\end{figure}

\paragraph{Bench holder}
We use an aluminium frame to hold the optical benches, which is depicted in Fig.~\ref{fig:lockeroverview} and~\ref{fig:ZerodurBenches}. 
This structure holds the sides of the optical benches. 
We use rubber between the optical bench and the holder to cushion vibrations and evenly distribute any forces from the holder.
For better thermal contact, we use heat-strips to connect the \acp{AOM} thermally to the aluminium frame.

A \ac{PCB} on top of the aluminium frame collects and multiplexes the electric signals from monitoring \acp{PD} and thermistors. It is also used for routing the electric signals for the shutter solenoids.

All optical benches are attached to a single metal plate, which sits at the side of the laser system locker. This plate also holds the spectroscopy benches, which are discussed in Sec.~\ref{sec:stab}.

\subsection{Fibre-optic based distribution} \label{sec:fibres}

In addition to the Zerodur boards, a number of commercially available fibre-based components will be used for light manipulation.
In the following section we will detail these components and their functions.

Fibre switches (FO Switch EOL PM from Weinert Industries) are used to switch between different light fields. 
The switches primarily allow us to switch the function of the Raman lasers.
Light from these lasers can either be used for interferometry, on one of two axes,
or it can be used on the detection axes, for either imaging or for quantum optics applications.

Fibre-based \acp{AOM} (fibre-Q by Gooch~\&~Housego) are used within the interferometry path. They are used to guarantee an optimal temporal overlap of the interferometry pulses for rubidium and potassium.

Additionally, there are a number of passive components in the distribution system, such as inline \acp{PD} from OZ optics.
These fibre based \acp{PD} allow light intensities to be monitored at different positions in the system without direct interaction with the hardware, forming a key part of the sophisticated housekeeping and monitoring system.

For the simple overlapping and splitting of light fields, especially for the 2D- and 3D-MOT, fixed ratio fibre splitters (954P by Evanescent Optics Inc.) are used. 
To this end, the light fields coming from one Zerodur board (usually containing light at both \SI{780}{\nano \meter} and \SI{767}{\nano \meter}) are superimposed with light coming from another Zerodur board. 
This allows for the generation of four light beams, each containing the four frequencies required for the 2D- and 3D-MOTs. 
To guarantee an optimal power ratio for the counter-propagating beams in the 3D-MOT and optical molasses, some of the fibre splitters are located inside the physics package locker.
This ensures that the power ratios are independent of the varying losses of multiple mating sleeves. 
Additionally, the configuration of the fibre splitters is chosen such that the wavelength dependent differences in splitting ratio are compensated for (see Fig.~\ref{fig:overview}).

\subsection{Frequency stabilization} \label{sec:stab}

\begin{figure}[ht!]
	\centering
	\includegraphics[width=\textwidth,height=\textheight,keepaspectratio]{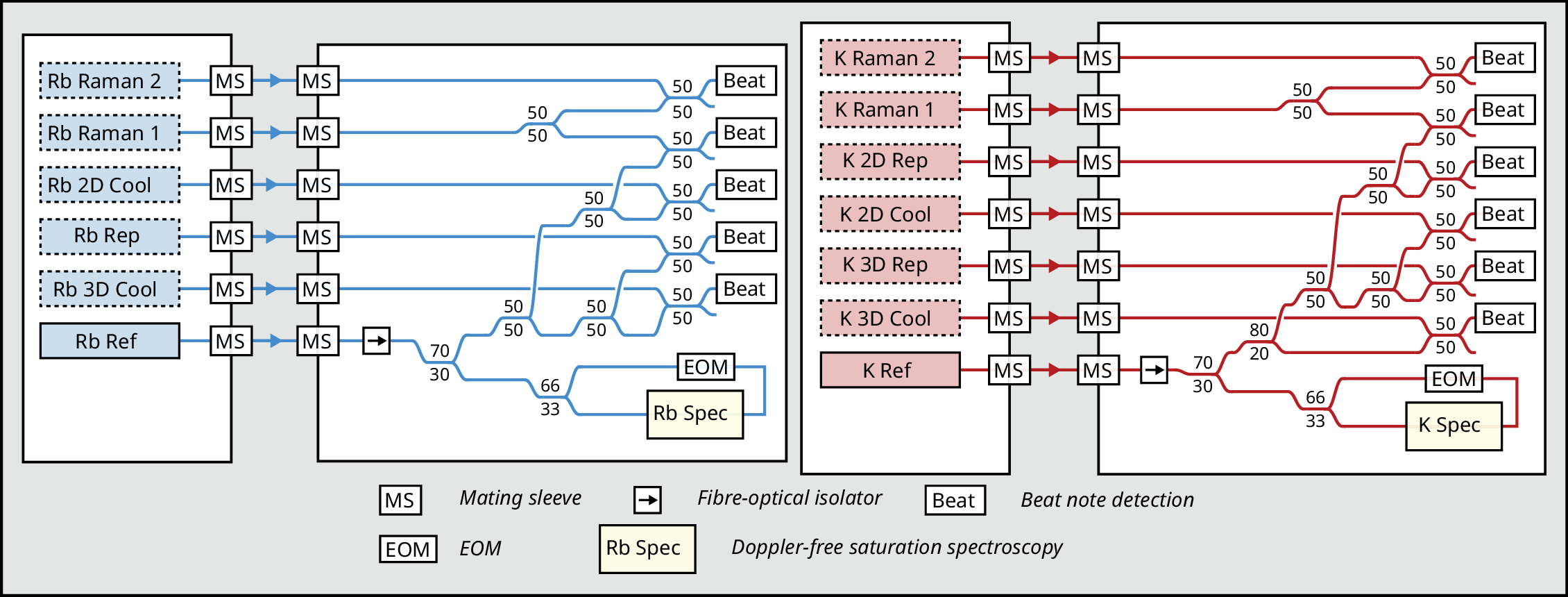}
	\caption{Optical schematic used for frequency stabilization. One laser for each species (named `Ref') is stabilized via Doppler-free saturation spectroscopy to an atomic transition. The other lasers, with the exception of the ``Raman 2" lasers, are stabilized via offset locking to these reference lasers. The ``Raman 2" lasers are stabilized relative to the ``Raman 1" lasers.}
	\label{fig:frequency_stab}
\end{figure}

We utilize several locking techniques in order to realize the required frequency stability, accuracy and tuning agility needed for the lasers at \SI{767}{\nano\meter} and \SI{780}{\nano\meter}.
Two lasers serve as reference lasers, one each at \SI{767}{\nano\metre} and \SI{780}{\nano\metre}, which are locked to atomic transitions of K and Rb, respectively. 
The remaining lasers are stabilized relative to the reference lasers via offset locking, using fast photoreceivers to create a beat-note. 
An exception to this are the ``Raman 2" lasers, which are used for atom interferometry.
These are stabilized relative to the ``Raman 1" lasers in a phase lock. 
One array of fibre splitters is used per species in order to create the necessary beat-notes.
The corresponding schematics can be seen in Fig.~\ref{fig:frequency_stab}

\begin{figure}[bt!]
	\centering
	\input{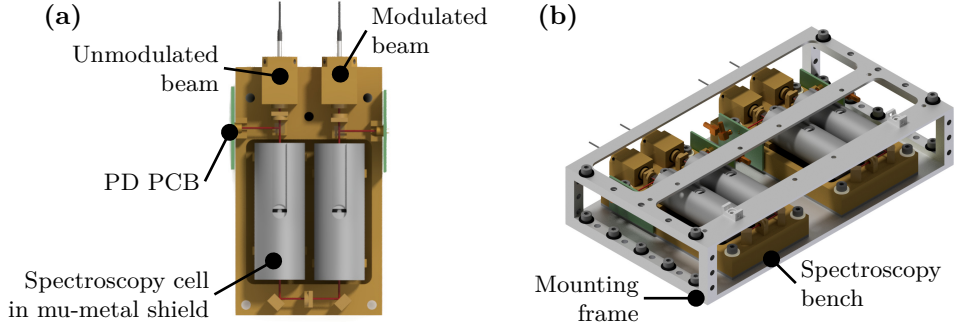}
	\caption{a) Rendering of a spectroscopy bench which is used for laser stablization with respect to an atomic transition. b) Rendering of the two spectroscopy benches integrated into their mounting frame. The benches will be screwed directly to the back wall of the mounting frame.}
	\label{fig:frequency_stab_bench}
\end{figure}

The locking of the reference lasers is achieved on two spectroscopy benches as shown in Fig.~\ref{fig:frequency_stab_bench}.
The optical fibre going to this bench is split into two, with one of the output fibres guided through a fibred \ac{EOM} to generate the sidebands necessary for Doppler-free \ac{FMS} and \ac{MTS}. 
Both beams enter the spectroscopy benches through collimators with focal lengths of $f=\SI{7.5}{\milli\metre}$. 
They are built using the same toolkit as the distribution benches.
The two beams are then guided in a counter-propagating fashion through two spectroscopy cells. 
Each beam is then guided onto a \ac{PD} to record the spectroscopy signal. 
This allows us to perform both \ac{FMS} and \ac{MTS}, which is required to enable stabilization to a multitude of spectroscopic transitions and also to gain from the superior frequency stability of the \ac{MTS} scheme.
The spectroscopy cells for potassium are heated to increase the vapour pressure and enhance the absorption signal.

\subsection{Intensity stabilization}

\begin{figure}[h!]
	\centering
	\includegraphics[width=0.9\textwidth, keepaspectratio]{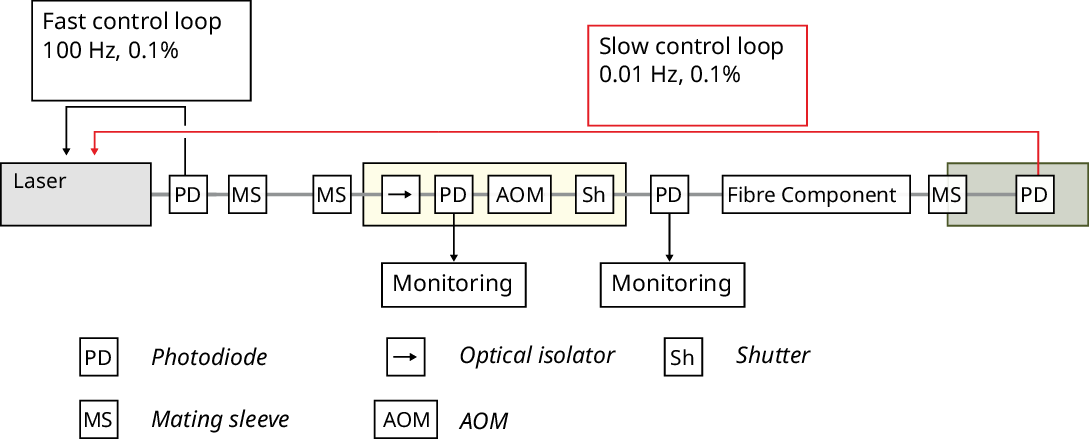}
	\caption{Schematic of the power control loop. A fast continuously running control loop will be used to stabilize the output power of the lasers. A slow control loop will be employed to compensate for fluctuations in the distribution system between experimental sequences.}
	\label{fig_int_stab}
\end{figure}

As \ac{BECCAL} sets very high requirements on the power stability of the light fields interacting with the atoms, the \ac{BECCAL} laser system will feature a distinct intensity stabilization system (see Fig.~\ref{fig_int_stab}). 
This system will operate in two ways. 
For fast intensity stabilization with bandwidths of at least \SI{100}{\hertz}, the light coupled into the fibre directly after the laser modules will be monitored and stabilized with a fast feedback loop to a predefined light intensity. 
Here the injection currents of the power amplifiers within the laser modules are used to control and stabilize the output power of the lasers.

As this control loop only compensates for fluctuations in the laser power and in the first fibre coupling, a second stabilization loop is also implemented. 
For this loop the light intensity at the final collimators in the physics package is measured at a specific time after each experimental sequence. 
This value will then be used to generate a global offset which will be applied to the injection currents of the power amplifiers to control the light intensities. 
This loop allows for the compensation of all effects attributed to the Zerodur boards, or the fibre based components, such as temperature dependent drifts on a shot-to-shot basis.

\section{Design of the \acs{COTS}-based ground system} \label{sec:GTB}

A second laser system, based entirely on \ac{COTS} parts, has been designed to meet the \ac{BECCAL} scientific requirements (Sec.~\ref{sec:Req}).
This system could be used to duplicate the \ac{BECCAL} \ac{FM} in a lab based environment and is planned to be used as one of the ground test-beds of the \ac{BECCAL} consortium.

The conceptual structure of the \ac{COTS}-based ground system closely resembles that of the \ac{FM}: the light from \ac{COTS} laser modules is distributed by fibre-based as well as free-space optics to deliver the required combinations of light fields to the physics package.
As it is built for lab-based experiments, it does not have to follow the strict \ac{SWaP} limitations and environmental constraints of the \ac{FM}, thus making it possible to use non-custom parts and reducing the complexity of laser replacement.
It means that not only are we not limited by space and vibration constraints or launch conditions, we can also rely on the system being located in a climate controlled facility.
Additionally, we do not require a dedicated \ac{ORU} concept for the \ac{COTS} model. 

\begin{figure}[ht]
	\centering
	\includegraphics[width=\textwidth]{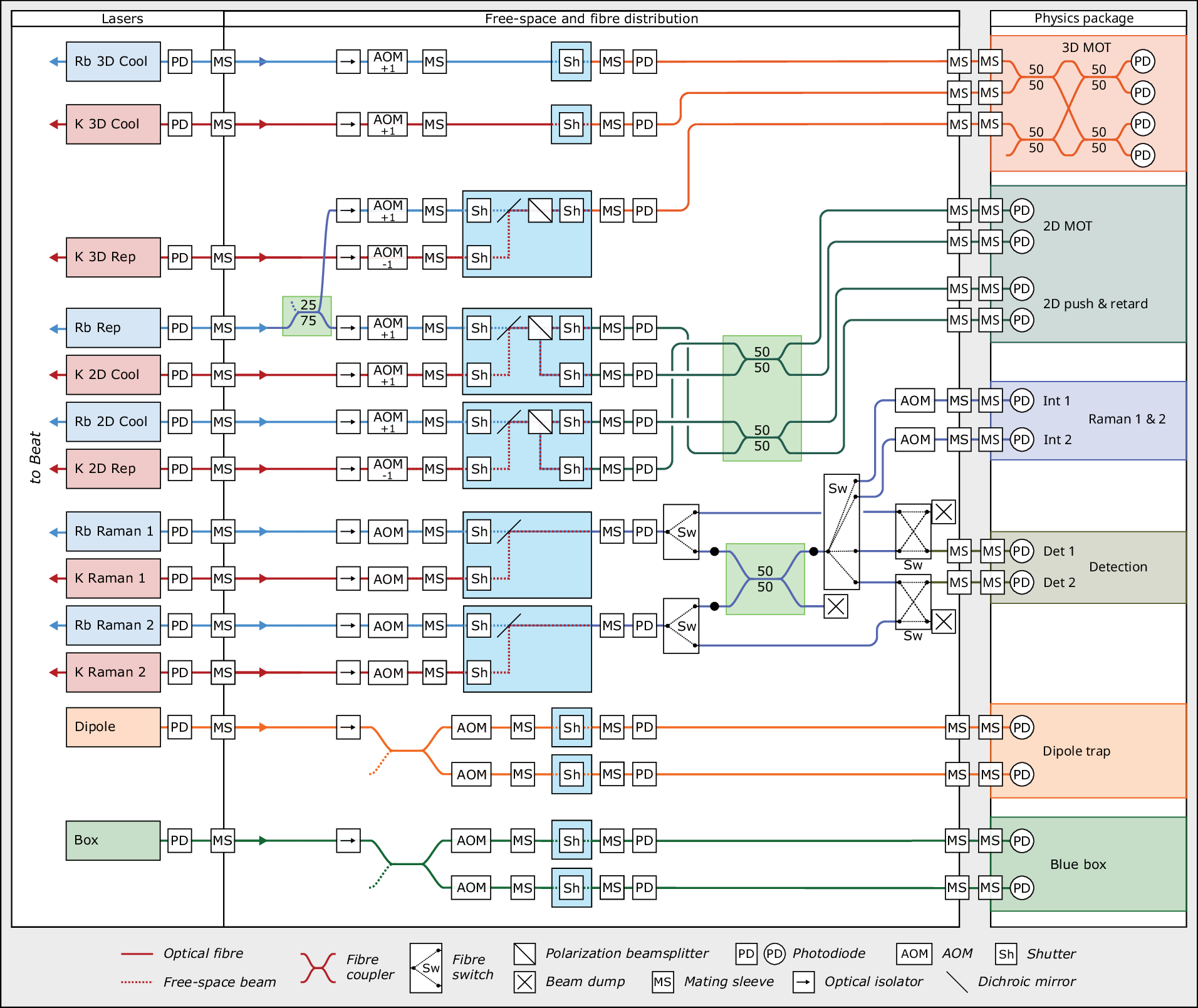}
	\caption{Optical schematic of the COTS-based laser system design. In the left panel the light of the \ac{COTS} lasers is picked up by inline photodiodes (\ac{PD}) for monitoring and stabilization and delivered to the free-space and fibre distribution section (central panel) by mating sleeves. Fibre-based optical isolators, \acp{AOM}, and switches are used as much as possible. Fibre port clusters (Sch\"after~+~Kirchhoff, shaded in light blue) host shutters and beam distribution optics that cannot effectively be matched by fibred components. The combined light fields are then delivered to the physics package (right panel) by another set of mating sleeves.}
	\label{fig:USGTBOverview}
\end{figure}

The priorities in designing this \ac{COTS} system are the commercial availability of components without additional development, similarity in control, behaviour, and interfaces to the flight model, and the ability to package for transfer between institutes.
Multiple \ac{COTS} options were considered for the different functional groups. Here we will detail the resulting design (Fig.~\ref{fig:USGTBOverview}).

\begin{figure}[ht]
	\centering
	\input{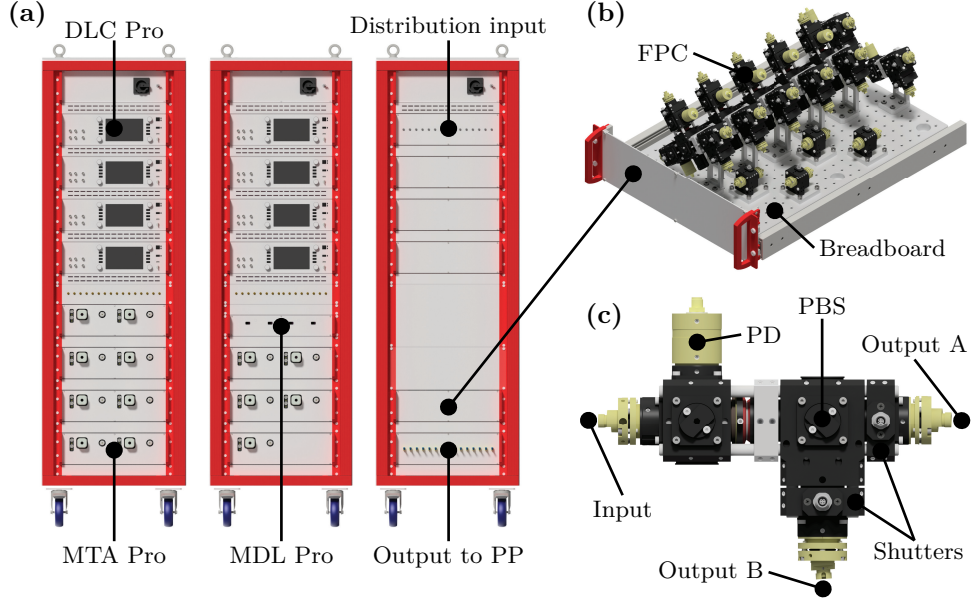}
	\caption{a) Rendering of the COTS-based ground system. The first two racks house thirteen amplified lasers (MTA Pros) and two unamplified lasers (MDL Pros) as well as their associated controllers. The third rack contains optical components for the distribution and manipulation of light fields. b) Rendering of a drawer from the third rack. Fibre port clusters (FPCs) are mounted on a breadboard for stability and accessibility. c) Rendering of a 1-to-2 fibre port cluster with integrated shutters and photodiode (PD).}
	\label{fig:GTBCADoverview}
\end{figure}

\subsection{Lasers}

As light sources we use MDL and MTA Pro Toptica lasers housed in 19 inch racks (T-Rack~\cite{tptc_BR-LRS-2021-04}, Fig.~\ref{fig:GTBCADoverview}).
Fifteen lasers provide the light fields for Rb and K 3D- and 2D-MOT cooling, dipole trapping, interferometry, and detection: 
six Rb lasers at \SI{780}{\nano\metre} (including one reference laser); seven K lasers at
\SI{767}{\nano\metre} (including one reference laser); one red-detuned dipole laser at \SI{1064}{\nano\metre}; one blue-detuned dipole laser at \SI{764}{\nano\metre}.
Utilizing \ac{COTS} components generally entails additional losses, as they cannot be optimized for transmission in their entirety as in the \ac{FM}.
These resulting higher losses have to be compensated for by the higher powers of the \ac{COTS} lasers compared to the \ac{FM} lasers.

\acp{ECDL} from Toptica (MDL Pro) are used as reference lasers at \SI{767}{\nano\metre} and \SI{780}{\nano\metre}~\cite{tptc_BR-LRS-2021-04}.
The remaining lasers in the system are an amplified version of this \ac{ECDL}: it is combined with an additional tapered amplifier module to form the Toptica MTA Pro~\cite{tptc_BR-LRS-2021-04}.
These lasers provide an optical output power, ex fibre, of \SIrange{50}{100}{\milli\watt} (MDL Pro) and \SIrange{800}{1000}{\milli\watt} (MTA Pro) at a typical power consumption of \SI{35}{\watt}. 
The polarization extinction ratio of light in the output fibre is greater than \SI{20}{\decibel}. 
Two laser modules come in one joint housing of $\SI{750}{\milli\metre}\! \times\! \SI{410}{\milli\metre}\! \times\! \SI{90}{\milli\metre}$ (including fibre couplers and strain reliefs) and a total mass of \SI{60}{\kilo\gram}.
Each set of two lasers is controlled by a 19 inch laser controller (DLC Pro) with a mass of \SI{8.5}{\kilo\gram}. 
In addition to the \ac{FM} laser functionality, the MTA Pro lasers have several extra capabilities.
Firstly, there is an automatic realignment function to optimize coupling of the seed laser into the tapered amplifier, as well as coupling into the output fibre.
Secondly, the laser drivers provide a coordinated electrical control which extends the mode-hop-free frequency tuning range further than is otherwise possible for this laser design.

\subsection{Free-space and fibred distribution}

The Zerodur benches are replaced by a combination of fibre port clusters (Sch\"after\,+\,Kirchhoff~\cite{suk_fibreprotcluster}) and fibred components.
The fibre port clusters are shaded in light blue in Fig.~\ref{fig:USGTBOverview}. 

These fibre port clusters, based on the Sch\"after\,+\,Kirchhoff multi-cube system with rectangular beam path geometry, combine fibre-optical free-space coupling with dichroic optics and shutters (48EMS-6, Sch\"after\,+\,Kirchhoff~\cite{suk_shutter}) to overlap and split the \SI{767}{\nano\metre} and \SI{780}{\nano\metre} laser light fields mirroring the Zerodur benches of the FM. 
In a typical fibre port cluster (like the one combining the repump fields for the Rb MOTs and the K 2D-MOT cooling light in Fig.~\ref{fig:USGTBOverview}), the light is first collimated by a beam coupler (60SMS-1-4). 
Bistable, electro-magnetic input shutters (48EMS-6) are used to selectively block either of the incoming \SI{767}{\nano\metre} or \SI{780}{\nano\metre} light fields.
A beam splitter (48BS) directs \SI{1}{\percent} of the optical power onto a \ac{PD} (4BPD-BPX61) to monitor coupling efficiencies throughout the system.
A waveplate adjusts the incoming fields into the accepted polarization angles of a dichroic beam combiner plate (48BC-CC).
Before the \ac{PBS} (48PM-S-B) that distributes the light to different output ports, two dichroic waveplates (48WP-2-780-1-767) are used to independently adjust the splitting ratios of the two wavelengths. 
More shutters are placed directly before the output fibre couplers (60SMS-1-4) and are used to fully extinguish a particular optical path.
The typical dimensions of one such fibre port cluster with two input and two output fibres are $\SI{400}{\milli\metre}\! \times\! \SI{300}{\milli\metre}\! \times\! \SI{90}{\milli\metre}$. 

In this arrangement, the fibre port clusters could not be combined with free-space \acp{AOM} and optical isolators as used on the Zerodur benches.
Instead we use fibred components, namely Thorlabs \acp{OI} (IO-J-780APC~\cite{thrl_oij780apc}), and Gooch\,\&\,Housego \acp{AOM} (S-M080-0.5C2W-3-F2P-01~\cite{gah_PEC0013iss8}).
These fibre components are deployed in the path prior to the fibre port clusters (Fig.~\ref{fig:USGTBOverview}).
The use of fibre based elements results in increased insertion losses of typically \SI{1.6}{\decibel} and \SI{2.5}{\decibel} compared to the free-space isolators and \acp{AOM} used in the FM respectively, this is due to the increase in the number of fibre to free-space interfaces.

The transmission efficiency of the fibre port clusters is mainly determined by the fibre coupling efficiency at the end of the free-space path.
This coupling efficiency is typically around \SI{65}{\percent}.
The overall transmission efficiency is expected to be approximately \SI{25}{\percent}.
One fibre port cluster including fibre isolators, \acp{AOM}, and fibre strain reliefs, which replicates the function of a 2-in-2-out Zerodur bench, takes a volume of $\SI{400}{\milli\metre}\! \times\! \SI{400}{\milli\metre}\! \times\! \SI{90}{\milli\metre}$ (\SI{14.4}{\liter}) and constitutes a mass of \SI{3}{\kilo\gram}. 

A similar approach to the \ac{FM} is taken for frequency stabilization. 
For the reference locks the flight model optical path is replicated as closely as possible by combining identical duplicates of the \ac{FM} spectroscopy cell with free-space \ac{COTS} optical components on a breadboard.
The design of the offset locks (Fig.~\ref{fig:frequency_stab}) is fully reproduced for the \ac{COTS} system, however, additional locking electronics from Toptica (DLC Pro Lock~\cite{tptc_prolock} and FALC Pro~\cite{tptc_falcpro}) can be utilized.

\section{Comparison of design approaches} \label{sec:Comp}

A direct comparison of the different and similar components in both systems is shown in Tab.~\ref{tab:comparison}.  
Here we see that both systems have common interfaces to other parts of the payload and also use the same \ac{COTS} fibre components, however the differences are the laser modules, and the free-space distribution, where custom-built items are replaced with the nearest \ac{COTS} equivalent regardless of size and weight.

We can further compare the two systems in Tabs.~\ref{tab:lasercomp} and~\ref{tab:ZDcomp}.  Here we see that the crucial properties of the two systems are comparable, and the major difference is optical power efficiency, size, weight, and electrical power consumption. 

\begin{table}[h]
	\caption{A comparison of the components used in the \ac{FM} and \ac{COTS}-based systems.}
	\label{tab:comparison}
	\begin{tabular*}{1\textwidth}{p{0.28\textwidth}p{0.32\textwidth}p{0.32\textwidth}}
		\toprule
		\textbf{Sub-part} & \textbf{\acs{FM}} & \textbf{\acs{COTS} system} \\ 
		\midrule
		&& \\
		Optical interfaces & \multicolumn{2}{c}{E2000} \\
		&& \\
		Laser modules & Custom \ac{ECDL}-\ac{MOPA} packages based on semi-conductor laser chips mounted on a microbench~\cite{Kurbis2020}. & Rack-mounted Toptica MTA modules including seeding laser and amplifier~\cite{tptc_BR-LRS-2021-04}. Two full laser-amplifier systems are integrated per rack drawer. \\
		&& \\
		Free-space distribution & Custom Zerodur ceramic benches with miniaturized optical components~\cite{Lezius2016,Dinkelaker2017,Becker2018a,schkolnik_compact_2016,mihm_zerodur_2019,mihm_phd}. & Customizable Sch\"after~+~Kirchhoff fibre port clusters~\cite{suk_fibreprotcluster}, fibred \acp{OI}~\cite{thrl_oij780apc} and fibred \acp{AOM}~\cite{gah_PEC0013iss8}. \\
		&& \\
		Fibre optic distribution & \multicolumn{2}{p{0.64\linewidth}}{Various \acs{COTS} parts including \acp{AOM} (Gooch\,\&\,Housego, Fibre-Q), splitters (Evanescent~Optics, 954P), switches (Weinert Fiber Optics, FO Switch EOL), and photodiodes (OZ Optics, OPM)} \\
		&& \\
		Frequency stabilization & Custom Zerodur ceramic ben- ches with miniaturized optical components~\cite{Lezius2016,Dinkelaker2017,Becker2018a,schkolnik_compact_2016,mihm_zerodur_2019,mihm_phd}, fibre splitters and \acs{COTS} electronics (Thorlabs, RX10CA) & \acs{COTS} optics and electronics including Toptica FALC Pro modules~\cite{tptc_falcpro}. \\
		&& \\
		Locker structure/housing & Custom-built locker to fit in \ac{EXPRESS} rack~\cite{Frye2021}. & Mounted in Toptica \num{19}'' racks \cite{tptc_BR-LRS-2021-04} with additional custom electrical patch panel. \\
		\botrule
	\end{tabular*}
\end{table}

The lasers used for the two systems both meet the laser system functionality requirements for \ac{BECCAL}.
The main difference between the two laser options is in terms of \ac{SWaP} budget.
The \ac{FM} lasers are $\SI{125}{\milli\meter}\! \times\! \SI{75}{\milli\meter}\! \times\! \SI{22.5}{\milli\meter}$ (\SI{0.2}{\liter}) and \SI{0.8}{\kilo\gram}, whereas each \ac{COTS} laser is nearly \num{40} times heavier and almost \num{70} times larger in volume at $\SI{750}{\milli\meter}\! \times\! \SI{205}{\milli\meter}\! \times\! \SI{90}{\milli\meter}$ (\SI{13.8}{\liter}) and  \SI{30}{\kilo\gram}.
Each \ac{COTS} laser also typically consumes \num{7} times more electrical power than an \ac{FM} laser.
The larger \ac{SWaP} budget allows the \ac{COTS} lasers to offer higher optical powers and better polarization extinction ratios.
This is due to a combination of factors.
When micro-fabricating lasers, one often needs to make a trade-off between component size and efficiency, this is particularly pertinent for components such as optical isolators.
The additional space in the \ac{COTS} lasers also allows for more flexibility in terms of alignment and additional optical components such as those used to control polarization.
The higher optical powers offered by the \ac{COTS} lasers are offset by higher losses in the \ac{COTS} distribution system.

\begin{table}
	\caption{Typical output parameters for the lasers used in both the \ac{FM} and \ac{COTS}-based systems. These parameters are taken from~\cite{Kurbis2020} for the \acs{FM} lasers, and from~\cite{tptc_linewidth, tptc_BR-LRS-2021-04} for the \acs{COTS} lasers.
	}
	\label{tab:lasercomp}
	\begin{tabular*}{\textwidth}{m{4cm}cccccc}
		\toprule
		& \multicolumn{3}{c}{\textbf{\acs{FM} lasers}} & \multicolumn{3}{c}{\textbf{\acs{COTS} lasers}} \\
		\cmidrule{2-4}\cmidrule{5-7}
		\textbf{Parameter} &  \textbf{\SI{767}{\nano\metre}} & \textbf{\SI{780}{\nano\metre}} & \textbf{\SI{1064}{\nano\metre}}  & \textbf{\SI{767}{\nano\metre}} & \textbf{\SI{780}{\nano\metre}} & \textbf{\SI{1064}{\nano\metre}} \\ 
		\midrule
		Output Power [\si{\milli\watt}]  & \num{250} & \num{300} & \num{500} & \num{800} & $\geq$~\num{1000} & \num{800} \\
		PER [dB] & \num{15} & \num{15} & \num{15} & $\geq$~\num{20} & $\geq$~\num{20} & $\geq$~\num{20} \\
		FWHM Linewidth [\si{\kilo\hertz}] & \multicolumn{3}{c}{$\leq$~\num{100} (\SI{1}{\milli\second})}  & \multicolumn{3}{c}{$\leq$~\num{100} (\SI{1}{\milli\second})} \\
		Mass per laser [\si{\kilo\gram}]& \multicolumn{3}{c}{$\approx$~\num{0.8}} & \multicolumn{3}{c}{$\approx$~\num{30}} \\
		Size per laser [\si{\milli\meter\cubed}]& \multicolumn{3}{c}{\num{125}\! $\times$\! \num{75}\! $\times$\! \num{22.5}} & \multicolumn{3}{c}{\num{750}\! $\times$\! \num{205}\! $\times$\! \num{90}} \\
		Power consumption per laser [\si{\watt}] & \multicolumn{3}{c}{$\leq$~\num{5}} & \multicolumn{3}{c}{typ. \num{35}} \\
		\botrule
	\end{tabular*}
\end{table}

The two different free-space distribution systems offer similar functionality in terms of properties such as switching speeds, but perform very differently when considering optical power efficiency and SWaP.

A single Zerodur bench, including shutters, \acp{AOM}, and isolators, has a maximum size of \SI{125}{\milli\meter}$\times$\SI{120}{\milli\meter}$\times$\SI{60}{\milli\meter} (or \SI{0.9}{\liter}) and \SI{1.2}{\kilogram}.
In comparison, a fibre port cluster plus fibred components equivalent to a single Zerodur bench occupies a total of \SI{400}{\milli\meter}$\times$\SI{400}{\milli\meter}$\times$\SI{90}{\milli\meter}(or \SI{14.4}{\liter}) and \SI{3}{\kilogram}.
This represents an approximately 16 fold increase in volume and a doubling of mass between the FM and \ac{COTS}-based systems.

The \ac{COTS}-based distribution system uses optical power almost half as efficiently than the micro-fabricated, miniaturized Zerodur benches.
This is because lower efficiency fibre optical \acp{AOM} and isolators are used in the \ac{COTS}-based system compared to the free-space components in the \ac{FM}.
In each of these fibre-based components, an additional fibre to free-space interface is present, introducing an additional source of losses.
Such an exchange is necessary as the free-space cube system cannot house the higher efficiency free-space components. 
We note additionally that the free-space components in the \ac{COTS}-based system may have a lower overall efficiency than the equivalent on the Zerodur benches due to differences in component choice and alignment techniques.

\begin{table}
	\caption{Comparison of expected laser system parameters including selected requirements. Typical efficiencies use the example path of 3D-MOT cooling light with the figures quoted a percentage of the light emitted from the laser diode or laser fibre which is then delivered to the physics package. We note that the mass, cost and size figures exclude electronics shared between the two systems, coolant, and connections to other systems.} 
	\label{tab:ZDcomp}
	\begin{tabular*}{\textwidth}{m{6cm}cc}
		\toprule
		\textbf{Parameter}  &  \textbf{\ac{FM}} & \textbf{\ac{COTS} system} \\
		\midrule
		Fast switching (via \ac{AOM}) & & \\
		- Switching Speed [\si{\nano\second}] &  \num{300} (\SIrange{0.05}{99.95}{\percent})  & \num{50} (\SIrange{10}{90}{\percent}) \\
		- Suppression Ratio  [\si{\decibel}] & $\leq \num{-30}$  & $\leq \num{-50}$  \\
		Slow switching (Shutter) & & \\
		- Switching Speed (\num{1} to \SI{99}{\percent}) [\si{\milli\second}] & \num{1} & \num{1.3}  \\
		- Suppression Ratio [\si{\decibel}] & $\leq \num{-120}$ & $\leq \num{-120}$ \\
		Typ.\ efficiency from diode output [\%] & \num{25(5)} & \num{12(3)} \\
		Typ.\ efficiency in distribution system [\%] & \num{35(5)} & \num{21(4)} \\
		Mass per Zerodur bench equivalent [\si{\kilo\gram}]& \num{1.2} & \num{3} \\
		Size per Zerodur bench equivalent [\si{\milli\metre^3}]& \num{125}\! $\times$\! \num{120}\! $\times$\! \num{60} & \num{400}\! $\times$\! \num{400}\! $\times$\! \num{90} \\
		\midrule
		Mass per system [\si{kg}]& \num{55} & $\leq$ \num{1200} \\
		Size per system [\si{\milli\metre^3}] & \num{445}\! $\times$\! \num{525}\! $\times$\! \num{270} & \num{800}\! $\times$\!\num{1800}\! $\times$\! \num{2050} \\
		Order of magnitude cost per system [\texteuro] & \num{4} million & \num{1.5} million \\
		\botrule
	\end{tabular*}
\end{table}

\begin{figure}
	\centering
	\includegraphics[width=\textwidth,height=\textheight,keepaspectratio]{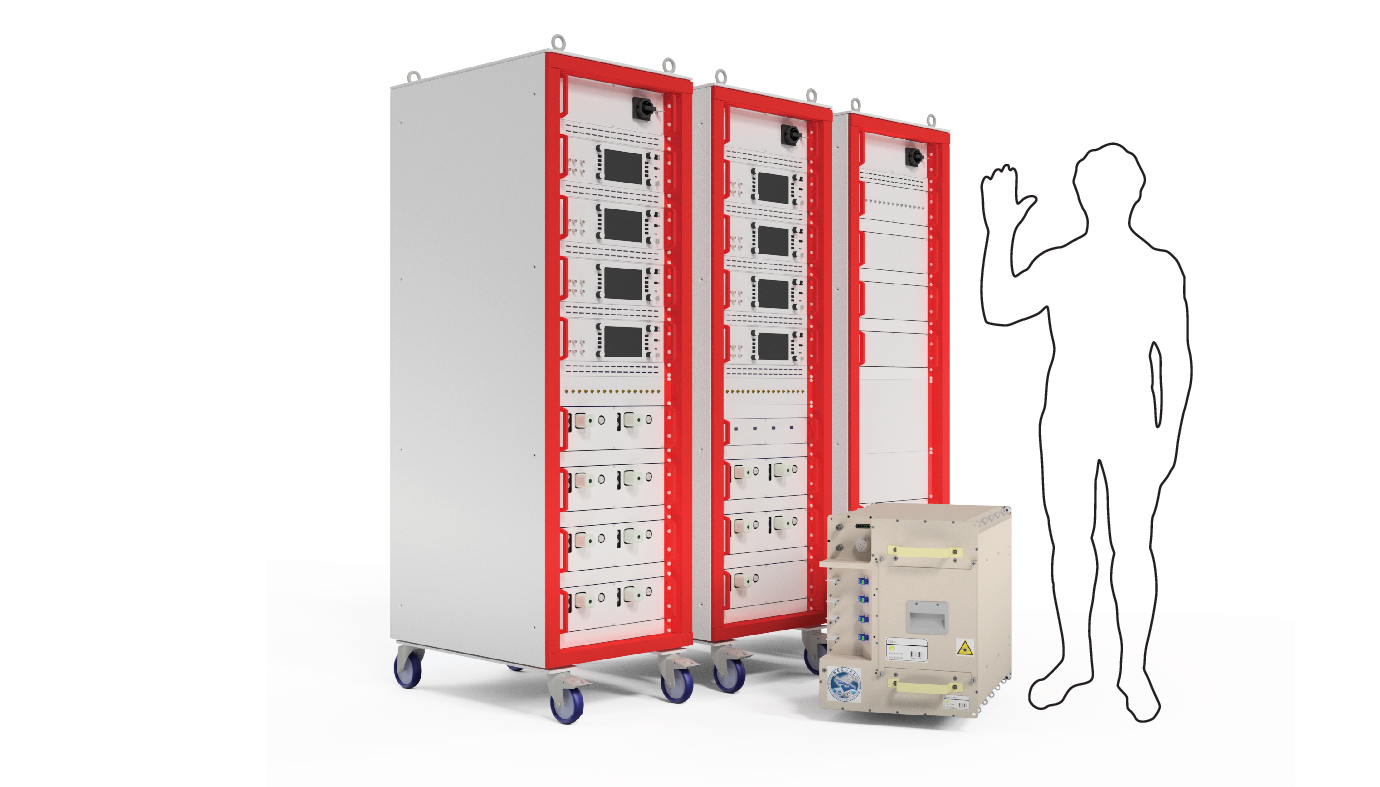}
	\caption[Size Comparison]{Rendering of the BECCAL flight model laser system locker alongside the larger COTS-based ground system. The height of each of the three COTS racks is \SI{1.97}{\metre} without eye bolts. The human for scale is adapted from the NASA Pioneer 10 plaque.}
	\label{fig:Size}
\end{figure}

As can be seen in Tab.~\ref{tab:lasercomp} and~\ref{tab:ZDcomp}, there is a significant difference between the \ac{SWaP} budgets of both systems.  
When combined with the locker housing (excluding cooling water, harness etc.), the \ac{FM} is \SI{445}{\milli\meter}$\times$\SI{525}{\milli\meter}$\times$\SI{270}{\milli\meter} (or \SI{63.1}{\liter}) and \SI{55}{\kilogram}.
In comparison, the \ac{COTS} ground system will fill three 19 inch racks, each with the dimensions \SI{800}{\milli\meter}$\times$\SI{600}{\milli\meter}$\times$\SI{2025}{\milli\meter}, 
 resulting in a total volume of \SI{2916}{\liter} or \SI{800}{\milli\meter}$\times$\SI{1800}{\milli\meter}$\times$\SI{2025}{\milli\meter}, and a mass of $\leq\SI{1200}{\kilogram}$ (max. \SI{400}{\kilogram} per T-Rack).
This is over forty times larger and up to twenty-two times heavier than the \ac{FM}, though we do note that the \ac{COTS}-based system does include the electronics required for laser control.
Each controller is \SI{8.5}{\kilo\gram} and thus a total of \SI{59.5}{\kilo\gram}.
Due to cooling requirements of the lasers, exclusion of the controllers does not lead to an overall volume reduction for the system. 
As the electronics for the \ac{FM} are treated separately, we cannot 
compare the electrical power consumption of both systems as a whole.  
A direct comparison of the sizes of the two systems, with a person for scale, is shown in Fig.~\ref{fig:Size}. 

In summary, both systems are designed to fulfil the scientific requirements of the \ac{BECCAL} payload, however, the \ac{COTS}-based system is significantly larger, heavier and less efficient than the \ac{FM} system.
These drawbacks are however unproblematic in a typical optics lab environment, as one does not have significant constraints on the \ac{SWaP} budget or stringent environmental constraints such as temperature or vibrational loads.
Additionally, the \ac{COTS}-based system is overall cheaper and quicker to build.

\section{Conclusion and outlook}

Two laser systems are presented which fulfil the functional requirements of the \ac{BECCAL} payload. One system also satisfies the environmental and \ac{SWaP} requirements of a flight system by utilising custom-built laser modules (\ac{ECDL}-\acp{MOPA}) and free-space optical benches made of Zerodur. The second is a cheaper and quicker but much larger and heavier \ac{COTS}-based system which is suitable for operation in a standard optics lab.

The designs presented here represent the current status of the laser system.  Integration of the \ac{COTS}-based system as a ground test-bed, and the first flight model system are currently in the initial phases and are expected to be delivered in the coming years. \ac{BECCAL} is expected to launch to the \ac{ISS} in 2027.

\section*{List of abbreviations}

\begin{acronym}
	\acro{AOM}{acousto-optic modulator}
	\acro{AR}{anti-reflection}
	\acro{BECCAL}{Bose-Einstein Condensate and Cold Atom Laboratory}
	\acro{CAL}{Cold Atom Lab}
	\acro{COTS}{commercial off the shelf}
	\acro{ECDL}{external-cavity diode laser}
	\acro{EOM}{electro-optic modulator}
	\acro{EXPRESS}{EXpedite the PRocessing of Experiments to the Space Station}
	\acro{fAOM}{fibre based \ac{AOM}}
	\acro{FM}{flight model}
	\acro{FMS}{frequency modulation spectroscopy}
	\acro{FPC}{fibre port cluster}
	\acro{FWHM}{full width at half maximum}
	\acro{ISS}{International Space Station}
	\acro{MOPA}{master oscillator power amplifier}
	\acro{MOT}{magneto-optical trap}
	\acro{MS}{mating sleeve}
	\acro{MTS}{modulation transfer spectroscopy}
	\acro{OI}{optical isolator}
	\acro{ORU}{orbital replacable unit}
	\acro{PBS}{polarizing beam splitters}
	\acro{PCB}{printed circuit board}
	\acro{PD}{photodiode}
	\acro{PP}{physics package}
	\acro{SWaP}{size, weight and power}	
	\acro{WP}{waveplate}
\end{acronym}

\backmatter

\section*{Availability of supporting data}
Not applicable

\section*{Competing interests}
The authors declare that they have no competing interests.

\section*{Funding}

The herein described project is a bilateral collaboration between NASA and DLR, both contributing to the scientific and operational organization. 

This work is supported by the German Space Agency (DLR) with funds provided by the Federal Ministry for Economic Affairs and Climate Action (BMWK) due to an enactment of the German Bundestag under Grant Nos. DLR50WP1431-1435, 50WM1131-1137, 50WM1141, 50WM1545, 50WM0937-0940, 50WM1237-1240, 50WP1700-1706, and 50WP2101-2104.

\section*{Author's contributions}

All authors read and approved the final manuscript.

\noindent V.A.H., J.P.M, A.We., T.K., and A.B. wrote the manuscript with additional support for figures from H.B. and M.K..

\noindent V.A.H., T.K., and J.P. had overall responsibility for the design and management of the laser systems led by A.P. and supported by the rest of the team at HU (H.B., M.K., C.G., E.V.K., B.L., H.T.S., C.W.).

\noindent The \ac{ECDL}-\ac{MOPA} modules are designed by the group of A.Wi. at FBH, with specific contributions relating to this paper given by A.B., M.D., C.K., A.Wi., supported by J.M.B., A.H., T.F., K.H\"{a}., M.S., R.S., and J.S..

\noindent The Zerodur benches are conceived, designed and manufactured by teams in Mainz and Hamburg including A.We. J.P.M, O.H., K.Se., S.B., E.dP.R, F.A.S., B.H., and P.W..

\noindent The overall architecture of the distribution and laser system are contributed to by V.A.H., J.P.M., A.We., T.K., M.K., J.P. with early work also completed by C.G..

\noindent M.W. and M.L.C. contributed to the design of the \ac{FM} infrastructure.

\noindent T.K., V.A.H, and H.B. designed the \ac{COTS}-based system.

\section*{Acknowledgements}

We in particular wish to acknowledge the contributions of the entire \ac{BECCAL} collaboration team.  The work described in this paper is merely a sub-set of a much wider and larger design and payload.
\ac{BECCAL} is a bilateral collaboration between NASA and DLR, both contributing to the scientific and operational organization.
The team acknowledges the contributions from NASA and their aid in adapting the payload to the needs of the International Space Station.

\bibliography{sources_reduced}


\begin{thebibliography}{45}
\ifx \bisbn   \undefined \def \bisbn  #1{ISBN #1}\fi
\ifx \binits  \undefined \def \binits#1{#1}\fi
\ifx \bauthor  \undefined \def \bauthor#1{#1}\fi
\ifx \batitle  \undefined \def \batitle#1{#1}\fi
\ifx \bjtitle  \undefined \def \bjtitle#1{#1}\fi
\ifx \bvolume  \undefined \def \bvolume#1{\textbf{#1}}\fi
\ifx \byear  \undefined \def \byear#1{#1}\fi
\ifx \bissue  \undefined \def \bissue#1{#1}\fi
\ifx \bfpage  \undefined \def \bfpage#1{#1}\fi
\ifx \blpage  \undefined \def \blpage #1{#1}\fi
\ifx \burl  \undefined \def \burl#1{\textsf{#1}}\fi
\ifx \doiurl  \undefined \def \doiurl#1{\url{https://doi.org/#1}}\fi
\ifx \betal  \undefined \def \betal{\textit{et al.}}\fi
\ifx \binstitute  \undefined \def \binstitute#1{#1}\fi
\ifx \binstitutionaled  \undefined \def \binstitutionaled#1{#1}\fi
\ifx \bctitle  \undefined \def \bctitle#1{#1}\fi
\ifx \beditor  \undefined \def \beditor#1{#1}\fi
\ifx \bpublisher  \undefined \def \bpublisher#1{#1}\fi
\ifx \bbtitle  \undefined \def \bbtitle#1{#1}\fi
\ifx \bedition  \undefined \def \bedition#1{#1}\fi
\ifx \bseriesno  \undefined \def \bseriesno#1{#1}\fi
\ifx \blocation  \undefined \def \blocation#1{#1}\fi
\ifx \bsertitle  \undefined \def \bsertitle#1{#1}\fi
\ifx \bsnm \undefined \def \bsnm#1{#1}\fi
\ifx \bsuffix \undefined \def \bsuffix#1{#1}\fi
\ifx \bparticle \undefined \def \bparticle#1{#1}\fi
\ifx \barticle \undefined \def \barticle#1{#1}\fi
\bibcommenthead
\ifx \bconfdate \undefined \def \bconfdate #1{#1}\fi
\ifx \botherref \undefined \def \botherref #1{#1}\fi
\ifx \url \undefined \def \url#1{\textsf{#1}}\fi
\ifx \bchapter \undefined \def \bchapter#1{#1}\fi
\ifx \bbook \undefined \def \bbook#1{#1}\fi
\ifx \bcomment \undefined \def \bcomment#1{#1}\fi
\ifx \oauthor \undefined \def \oauthor#1{#1}\fi
\ifx \citeauthoryear \undefined \def \citeauthoryear#1{#1}\fi
\ifx \endbibitem  \undefined \def \endbibitem {}\fi
\ifx \bconflocation  \undefined \def \bconflocation#1{#1}\fi
\ifx \arxivurl  \undefined \def \arxivurl#1{\textsf{#1}}\fi
\csname PreBibitemsHook\endcsname

\bibitem[\protect\citeauthoryear{Bongs et~al.}{2019}]{Bongs2019}
\begin{barticle}
\bauthor{\bsnm{Bongs}, \binits{K.}},
\bauthor{\bsnm{Holynski}, \binits{M.}},
\bauthor{\bsnm{Vovrosh}, \binits{J.}},
\bauthor{\bsnm{Bouyer}, \binits{P.}},
\bauthor{\bsnm{Condon}, \binits{G.}},
\bauthor{\bsnm{Rasel}, \binits{E.}}, \betal:
\batitle{{Taking atom interferometric quantum sensors from the laboratory to
  real-world applications}}.
\bjtitle{Nature Reviews Physics}
\bvolume{1}(\bissue{12}),
\bfpage{731}--\blpage{739}
(\byear{2019})
\doiurl{10.1038/s42254-019-0117-4}
\end{barticle}
\endbibitem

\bibitem[\protect\citeauthoryear{Alonso et~al.}{2022}]{Alonso2022}
\begin{barticle}
\bauthor{\bsnm{Alonso}, \binits{I.}},
\bauthor{\bsnm{Alpigiani}, \binits{C.}},
\bauthor{\bsnm{Altschul}, \binits{B.}},
\bauthor{\bsnm{Ara{\'{u}}jo}, \binits{H.}},
\bauthor{\bsnm{Arduini}, \binits{G.}},
\bauthor{\bsnm{Arlt}, \binits{J.}}, \betal:
\batitle{{Cold atoms in space: community workshop summary and proposed
  road-map}}.
\bjtitle{EPJ Quantum Technology}
\bvolume{9}(\bissue{1}),
\bfpage{30}
(\byear{2022})
\doiurl{10.1140/epjqt/s40507-022-00147-w}
{\href{https://arxiv.org/abs/2201.07789}{{2201.07789}}}
\end{barticle}
\endbibitem

\bibitem[\protect\citeauthoryear{Abend et~al.}{2023}]{Abend2023}
\begin{botherref}
\oauthor{\bsnm{Abend}, \binits{S.}},
\oauthor{\bsnm{Allard}, \binits{B.}},
\oauthor{\bsnm{Arnold}, \binits{A.S.}},
\oauthor{\bsnm{Ban}, \binits{T.}},
\oauthor{\bsnm{Barry}, \binits{L.}},
\oauthor{\bsnm{Battelier}, \binits{B.}}, et al.:
{Technology roadmap for cold-atoms based quantum inertial sensor in space}.
AVS Quantum Science
\textbf{5}(1)
(2023)
\doiurl{10.1116/5.0098119}
\end{botherref}
\endbibitem

\bibitem[\protect\citeauthoryear{Frye et~al.}{}]{Frye2021}
\begin{botherref}
\oauthor{\bsnm{Frye}, \binits{K.}},
\oauthor{\bsnm{Abend}, \binits{S.}},
\oauthor{\bsnm{Bartosch}, \binits{W.}},
\oauthor{\bsnm{Bawamia}, \binits{A.}},
\oauthor{\bsnm{Becker}, \binits{D.}},
\oauthor{\bsnm{Blume}, \binits{H.}}, et al.:
{The Bose-Einstein Condensate and Cold Atom Laboratory}
\textbf{8}(1),
1--38
\doiurl{10.1140/epjqt/s40507-020-00090-8}
{\href{https://arxiv.org/abs/1912.04849}{{1912.04849}}}
\end{botherref}
\endbibitem

\bibitem[\protect\citeauthoryear{Sabulsky et~al.}{2020}]{Sabulsky2020}
\begin{barticle}
\bauthor{\bsnm{Sabulsky}, \binits{D.O.}},
\bauthor{\bsnm{Junca}, \binits{J.}},
\bauthor{\bsnm{Lef{\`{e}}vre}, \binits{G.}},
\bauthor{\bsnm{Zou}, \binits{X.}},
\bauthor{\bsnm{Bertoldi}, \binits{A.}},
\bauthor{\bsnm{Battelier}, \binits{B.}}, \betal:
\batitle{{A fibered laser system for the MIGA large scale atom
  interferometer}}.
\bjtitle{Scientific Reports}
\bvolume{10}(\bissue{1}),
\bfpage{1}--\blpage{16}
(\byear{2020})
\doiurl{10.1038/s41598-020-59971-8}
{\href{https://arxiv.org/abs/1911.12209}{{1911.12209}}}
\end{barticle}
\endbibitem

\bibitem[\protect\citeauthoryear{Canuel et~al.}{2022}]{Canuel2022}
\begin{botherref}
\oauthor{\bsnm{Canuel}, \binits{B.}},
\oauthor{\bsnm{Zou}, \binits{X.}},
\oauthor{\bsnm{Sabulsky}, \binits{D.O.}},
\oauthor{\bsnm{Junca}, \binits{J.}},
\oauthor{\bsnm{Bertoldi}, \binits{A.}},
\oauthor{\bsnm{Beaufils}, \binits{Q.}}, et al.:
{A gravity antenna based on quantum technologies: MIGA},
6--9
(2022)
{\href{https://arxiv.org/abs/2204.12137}{{arXiv:2204.12137}}}
\end{botherref}
\endbibitem

\bibitem[\protect\citeauthoryear{Badurina et~al.}{2020}]{Badurina2020}
\begin{botherref}
\oauthor{\bsnm{Badurina}, \binits{L.}},
\oauthor{\bsnm{Bentine}, \binits{E.}},
\oauthor{\bsnm{Blas}, \binits{D.}},
\oauthor{\bsnm{Bongs}, \binits{K.}},
\oauthor{\bsnm{Bortoletto}, \binits{D.}},
\oauthor{\bsnm{Bowcock}, \binits{T.}}, et al.:
{AION: An atom interferometer observatory and network}.
Journal of Cosmology and Astroparticle Physics
\textbf{2020}(5)
(2020)
\doiurl{10.1088/1475-7516/2020/05/011}
{\href{https://arxiv.org/abs/1911.11755}{{1911.11755}}}
\end{botherref}
\endbibitem

\bibitem[\protect\citeauthoryear{Zhan et~al.}{2020}]{Zhan2020}
\begin{barticle}
\bauthor{\bsnm{Zhan}, \binits{M.S.}},
\bauthor{\bsnm{Wang}, \binits{J.}},
\bauthor{\bsnm{Ni}, \binits{W.T.}},
\bauthor{\bsnm{Gao}, \binits{D.F.}},
\bauthor{\bsnm{Wang}, \binits{G.}},
\bauthor{\bsnm{He}, \binits{L.X.}}, \betal:
\batitle{{ZAIGA: Zhaoshan long-baseline atom interferometer gravitation
  antenna}}.
\bjtitle{International Journal of Modern Physics D}
\bvolume{29}(\bissue{4}),
\bfpage{1}--\blpage{20}
(\byear{2020})
\doiurl{10.1142/S0218271819400054}
{\href{https://arxiv.org/abs/1903.09288}{{1903.09288}}}
\end{barticle}
\endbibitem

\bibitem[\protect\citeauthoryear{Canuel et~al.}{2020}]{Canuel2020}
\begin{botherref}
\oauthor{\bsnm{Canuel}, \binits{B.}},
\oauthor{\bsnm{Abend}, \binits{S.}},
\oauthor{\bsnm{Amaro-Seoane}, \binits{P.}},
\oauthor{\bsnm{Badaracco}, \binits{F.}},
\oauthor{\bsnm{Beaufils}, \binits{Q.}},
\oauthor{\bsnm{Bertoldi}, \binits{A.}}, et al.:
{Technologies for the ELGAR large scale atom interferometer array}
(2020)
{\href{https://arxiv.org/abs/2007.04014}{{arXiv:2007.04014}}}
\end{botherref}
\endbibitem

\bibitem[\protect\citeauthoryear{El-Neaj et~al.}{2020}]{El-Neaj2020}
\begin{botherref}
\oauthor{\bsnm{El-Neaj}, \binits{Y.A.}},
\oauthor{\bsnm{Alpigiani}, \binits{C.}},
\oauthor{\bsnm{Amairi-Pyka}, \binits{S.}},
\oauthor{\bsnm{Ara{\'{u}}jo}, \binits{H.}},
\oauthor{\bsnm{Bala{\v{z}}}, \binits{A.}},
\oauthor{\bsnm{Bassi}, \binits{A.}}, et al.:
{AEDGE: Atomic Experiment for Dark Matter and Gravity Exploration in Space}.
EPJ Quantum Technology
\textbf{7}(1)
(2020)
\doiurl{10.1140/epjqt/s40507-020-0080-0}
\end{botherref}
\endbibitem

\bibitem[\protect\citeauthoryear{L{\'{e}}v{\`{e}}que
  et~al.}{2022}]{Leveque2022}
\begin{botherref}
\oauthor{\bsnm{L{\'{e}}v{\`{e}}que}, \binits{T.}},
\oauthor{\bsnm{Fallet}, \binits{C.}},
\oauthor{\bsnm{Lefebve}, \binits{J.}},
\oauthor{\bsnm{Piquereau}, \binits{A.}},
\oauthor{\bsnm{Gauguet}, \binits{A.}},
\oauthor{\bsnm{Battelier}, \binits{B.}}, et al.:
{CARIOQA: Definition of a Quantum Pathfinder Mission}
(July)
(2022)
\doiurl{10.1117/12.2690536}
{\href{https://arxiv.org/abs/2211.01215}{{2211.01215}}}
\end{botherref}
\endbibitem

\bibitem[\protect\citeauthoryear{Ahlers et~al.}{2022}]{Ahlers2022}
\begin{botherref}
\oauthor{\bsnm{Ahlers}, \binits{H.}},
\oauthor{\bsnm{Badurina}, \binits{L.}},
\oauthor{\bsnm{Bassi}, \binits{A.}},
\oauthor{\bsnm{Battelier}, \binits{B.}},
\oauthor{\bsnm{Beaufils}, \binits{Q.}},
\oauthor{\bsnm{Bongs}, \binits{K.}}, et al.:
{STE-QUEST: Space Time Explorer and QUantum Equivalence principle Space Test}
(2022)
{\href{https://arxiv.org/abs/2211.15412}{{arXiv:2211.15412}}}
\end{botherref}
\endbibitem

\bibitem[\protect\citeauthoryear{Thompson et~al.}{2022}]{Thompson2022}
\begin{barticle}
\bauthor{\bsnm{Thompson}, \binits{R.J.}},
\bauthor{\bsnm{Aveline}, \binits{D.C.}},
\bauthor{\bsnm{Chiow}, \binits{S.-W.}},
\bauthor{\bsnm{Elliott}, \binits{E.R.}},
\bauthor{\bsnm{Kellogg}, \binits{J.R.}},
\bauthor{\bsnm{Kohel}, \binits{J.M.}}, \betal:
\batitle{{Exploring the quantum world with a third generation Ultra-cold atom
  facility}}.
\bjtitle{Quantum Science and Technology}
\bvolume{29}(\bissue{46}),
\bfpage{465705}
(\byear{2022})
\doiurl{10.1088/2058-9565/aca34f}
\end{barticle}
\endbibitem

\bibitem[\protect\citeauthoryear{Aveline et~al.}{2020}]{Aveline2020}
\begin{barticle}
\bauthor{\bsnm{Aveline}, \binits{D.C.}},
\bauthor{\bsnm{Williams}, \binits{J.R.}},
\bauthor{\bsnm{Elliott}, \binits{E.R.}},
\bauthor{\bsnm{Dutenhoffer}, \binits{C.}},
\bauthor{\bsnm{Kellogg}, \binits{J.R.}},
\bauthor{\bsnm{Kohel}, \binits{J.M.}}, \betal:
\batitle{{Observation of Bose–Einstein condensates in an Earth-orbiting
  research lab}}.
\bjtitle{Nature}
\bvolume{582}(\bissue{7811}),
\bfpage{193}--\blpage{197}
(\byear{2020})
\doiurl{10.1038/s41586-020-2346-1}
\end{barticle}
\endbibitem

\bibitem[\protect\citeauthoryear{Liu et~al.}{2018}]{Liu2018}
\begin{barticle}
\bauthor{\bsnm{Liu}, \binits{L.}},
\bauthor{\bsnm{L{\"{u}}}, \binits{D.-S.}},
\bauthor{\bsnm{Chen}, \binits{W.-B.}},
\bauthor{\bsnm{Li}, \binits{T.}},
\bauthor{\bsnm{Qu}, \binits{Q.-Z.}},
\bauthor{\bsnm{Wang}, \binits{B.}}, \betal:
\batitle{{In-orbit operation of an atomic clock based on laser-cooled 87Rb
  atoms}}.
\bjtitle{Nature Communications}
\bvolume{9}(\bissue{1}),
\bfpage{2760}
(\byear{2018})
\doiurl{10.1038/s41467-018-05219-z}
\end{barticle}
\endbibitem

\bibitem[\protect\citeauthoryear{Laurent et~al.}{2015}]{Laurent2015}
\begin{barticle}
\bauthor{\bsnm{Laurent}, \binits{P.}},
\bauthor{\bsnm{Massonnet}, \binits{D.}},
\bauthor{\bsnm{Cacciapuoti}, \binits{L.}},
\bauthor{\bsnm{Salomon}, \binits{C.}}:
\batitle{{The ACES/PHARAO space mission}}.
\bjtitle{Comptes Rendus Physique}
\bvolume{16}(\bissue{5}),
\bfpage{540}--\blpage{552}
(\byear{2015})
\doiurl{10.1016/j.crhy.2015.05.002}
\end{barticle}
\endbibitem

\bibitem[\protect\citeauthoryear{Li et~al.}{2023}]{Li2023}
\begin{barticle}
\bauthor{\bsnm{Li}, \binits{L.}},
\bauthor{\bsnm{Xiong}, \binits{W.}},
\bauthor{\bsnm{Wang}, \binits{B.}},
\bauthor{\bsnm{Li}, \binits{T.}},
\bauthor{\bsnm{Xie}, \binits{Y.}},
\bauthor{\bsnm{Liang}, \binits{A.}}, \betal:
\batitle{{The Design, Realization, and Validation of the Scheme for Quantum
  Degenerate Research in Microgravity}}.
\bjtitle{IEEE Photonics Journal}
\bvolume{15}(\bissue{3}),
\bfpage{1}--\blpage{8}
(\byear{2023})
\doiurl{10.1109/JPHOT.2023.3266108}
\end{barticle}
\endbibitem

\bibitem[\protect\citeauthoryear{Schkolnik et~al.}{2016}]{Schkolnik2016}
\begin{barticle}
\bauthor{\bsnm{Schkolnik}, \binits{V.}},
\bauthor{\bsnm{Hellmig}, \binits{O.}},
\bauthor{\bsnm{Wenzlawski}, \binits{A.}},
\bauthor{\bsnm{Grosse}, \binits{J.}},
\bauthor{\bsnm{Kohfeldt}, \binits{A.}},
\bauthor{\bsnm{D{\"{o}}ringshoff}, \binits{K.}}, \betal:
\batitle{{A compact and robust diode laser system for atom interferometry on a
  sounding rocket}}.
\bjtitle{Applied Physics B}
\bvolume{122}(\bissue{8}),
\bfpage{217}
(\byear{2016})
\doiurl{10.1007/s00340-016-6490-0}
\end{barticle}
\endbibitem

\bibitem[\protect\citeauthoryear{Becker et~al.}{}]{Becker2018a}
\begin{botherref}
\oauthor{\bsnm{Becker}, \binits{D.}},
\oauthor{\bsnm{Lachmann}, \binits{M.D.}},
\oauthor{\bsnm{Seidel}, \binits{S.T.}},
\oauthor{\bsnm{Ahlers}, \binits{H.}},
\oauthor{\bsnm{Dinkelaker}, \binits{A.N.}},
\oauthor{\bsnm{Grosse}, \binits{J.}}, et al.:
Space-borne bose–einstein condensation for precision interferometry
\textbf{562}(7727),
391--395
\doiurl{10.1038/s41586-018-0605-1}
\end{botherref}
\endbibitem

\bibitem[\protect\citeauthoryear{D{\"{o}}ringshoff
  et~al.}{2019}]{Doringshoff2019}
\begin{barticle}
\bauthor{\bsnm{D{\"{o}}ringshoff}, \binits{K.}},
\bauthor{\bsnm{Gutsch}, \binits{F.B.}},
\bauthor{\bsnm{Schkolnik}, \binits{V.}},
\bauthor{\bsnm{K{\"{u}}rbis}, \binits{C.}},
\bauthor{\bsnm{Oswald}, \binits{M.}},
\bauthor{\bsnm{Pr{\"{o}}bster}, \binits{B.}}, \betal:
\batitle{{Iodine Frequency Reference on a Sounding Rocket}}.
\bjtitle{Physical Review Applied}
\bvolume{11}(\bissue{5}),
\bfpage{054068}
(\byear{2019})
\doiurl{10.1103/PhysRevApplied.11.054068}
\end{barticle}
\endbibitem

\bibitem[\protect\citeauthoryear{Schkolnik et~al.}{2017}]{Schkolnik2017}
\begin{barticle}
\bauthor{\bsnm{Schkolnik}, \binits{V.}},
\bauthor{\bsnm{D{\"{o}}ringshoff}, \binits{K.}},
\bauthor{\bsnm{Gutsch}, \binits{F.B.}},
\bauthor{\bsnm{Oswald}, \binits{M.}},
\bauthor{\bsnm{Schuldt}, \binits{T.}},
\bauthor{\bsnm{Braxmaier}, \binits{C.}}, \betal:
\batitle{{JOKARUS - Design of a compact optical iodine frequency reference for
  a sounding rocket mission}}.
\bjtitle{EPJ Quantum Technology}
\bvolume{4}(\bissue{1}),
\bfpage{9}
(\byear{2017})
\doiurl{10.1140/epjqt/s40507-017-0063-y}
{\href{https://arxiv.org/abs/1702.08330}{{1702.08330}}}
\end{barticle}
\endbibitem

\bibitem[\protect\citeauthoryear{Lezius et~al.}{2016}]{Lezius2016}
\begin{barticle}
\bauthor{\bsnm{Lezius}, \binits{M.}},
\bauthor{\bsnm{Wilken}, \binits{T.}},
\bauthor{\bsnm{Deutsch}, \binits{C.}},
\bauthor{\bsnm{Giunta}, \binits{M.}},
\bauthor{\bsnm{Mandel}, \binits{O.}},
\bauthor{\bsnm{Thaller}, \binits{A.}}, \betal:
\batitle{{Space-borne frequency comb metrology}}.
\bjtitle{Optica}
\bvolume{3}(\bissue{12}),
\bfpage{1381}
(\byear{2016})
\doiurl{10.1364/OPTICA.3.001381}
\end{barticle}
\endbibitem

\bibitem[\protect\citeauthoryear{Dinkelaker et~al.}{2017}]{Dinkelaker2017}
\begin{barticle}
\bauthor{\bsnm{Dinkelaker}, \binits{A.N.}},
\bauthor{\bsnm{Schiemangk}, \binits{M.}},
\bauthor{\bsnm{Schkolnik}, \binits{V.}},
\bauthor{\bsnm{Kenyon}, \binits{A.}},
\bauthor{\bsnm{Lampmann}, \binits{K.}},
\bauthor{\bsnm{Wenzlawski}, \binits{A.}}, \betal:
\batitle{{Autonomous frequency stabilization of two extended-cavity diode
  lasers at the potassium wavelength on a sounding rocket}}.
\bjtitle{Applied Optics}
\bvolume{56}(\bissue{5}),
\bfpage{1388}
(\byear{2017})
\doiurl{10.1364/AO.56.001388}
\end{barticle}
\endbibitem

\bibitem[\protect\citeauthoryear{Rudolph et~al.}{2015}]{Rudolph2015}
\begin{barticle}
\bauthor{\bsnm{Rudolph}, \binits{J.}},
\bauthor{\bsnm{Herr}, \binits{W.}},
\bauthor{\bsnm{Grzeschik}, \binits{C.}},
\bauthor{\bsnm{Sternke}, \binits{T.}},
\bauthor{\bsnm{Grote}, \binits{A.}},
\bauthor{\bsnm{Popp}, \binits{M.}}, \betal:
\batitle{{A high-flux BEC source for mobile atom interferometers}}.
\bjtitle{New Journal of Physics}
\bvolume{17}(\bissue{6}),
\bfpage{065001}
(\byear{2015})
\doiurl{10.1088/1367-2630/17/6/065001}
\end{barticle}
\endbibitem

\bibitem[\protect\citeauthoryear{Deppner et~al.}{2021}]{Deppner2021}
\begin{barticle}
\bauthor{\bsnm{Deppner}, \binits{C.}},
\bauthor{\bsnm{Herr}, \binits{W.}},
\bauthor{\bsnm{Cornelius}, \binits{M.}},
\bauthor{\bsnm{Stromberger}, \binits{P.}},
\bauthor{\bsnm{Sternke}, \binits{T.}},
\bauthor{\bsnm{Grzeschik}, \binits{C.}}, \betal:
\batitle{{Collective-Mode Enhanced Matter-Wave Optics}}.
\bjtitle{Physical Review Letters}
\bvolume{127}(\bissue{10}),
\bfpage{100401}
(\byear{2021})
\doiurl{10.1103/PhysRevLett.127.100401}
\end{barticle}
\endbibitem

\bibitem[\protect\citeauthoryear{Vogt et~al.}{2020}]{Vogt2020}
\begin{barticle}
\bauthor{\bsnm{Vogt}, \binits{C.}},
\bauthor{\bsnm{Woltmann}, \binits{M.}},
\bauthor{\bsnm{Herrmann}, \binits{S.}},
\bauthor{\bsnm{L{\"{a}}mmerzahl}, \binits{C.}},
\bauthor{\bsnm{Albers}, \binits{H.}},
\bauthor{\bsnm{Schlippert}, \binits{D.}},
\bauthor{\bsnm{Rasel}, \binits{E.M.}}:
\batitle{{Evaporative cooling from an optical dipole trap in microgravity}}.
\bjtitle{Physical Review A}
\bvolume{101}(\bissue{1}),
\bfpage{1}--\blpage{6}
(\byear{2020})
\doiurl{10.1103/PhysRevA.101.013634}
{\href{https://arxiv.org/abs/1909.03800}{{1909.03800}}}
\end{barticle}
\endbibitem

\bibitem[\protect\citeauthoryear{Barrett et~al.}{2016}]{Barrett2016}
\begin{barticle}
\bauthor{\bsnm{Barrett}, \binits{B.}},
\bauthor{\bsnm{Antoni-Micollier}, \binits{L.}},
\bauthor{\bsnm{Chichet}, \binits{L.}},
\bauthor{\bsnm{Battelier}, \binits{B.}},
\bauthor{\bsnm{L{\'{e}}v{\`{e}}que}, \binits{T.}},
\bauthor{\bsnm{Landragin}, \binits{A.}},
\bauthor{\bsnm{Bouyer}, \binits{P.}}:
\batitle{{Dual matter-wave inertial sensors in weightlessness}}.
\bjtitle{Nature Communications}
\bvolume{7}(\bissue{1}),
\bfpage{13786}
(\byear{2016})
\doiurl{10.1038/ncomms13786}
\end{barticle}
\endbibitem

\bibitem[\protect\citeauthoryear{Condon et~al.}{2019}]{Condon2019}
\begin{barticle}
\bauthor{\bsnm{Condon}, \binits{G.}},
\bauthor{\bsnm{Rabault}, \binits{M.}},
\bauthor{\bsnm{Barrett}, \binits{B.}},
\bauthor{\bsnm{Chichet}, \binits{L.}},
\bauthor{\bsnm{Arguel}, \binits{R.}},
\bauthor{\bsnm{Eneriz-Imaz}, \binits{H.}}, \betal:
\batitle{{All-Optical Bose-Einstein Condensates in Microgravity}}.
\bjtitle{Physical Review Letters}
\bvolume{123}(\bissue{24}),
\bfpage{240402}
(\byear{2019})
\doiurl{10.1103/PhysRevLett.123.240402}
\end{barticle}
\endbibitem

\bibitem[\protect\citeauthoryear{Raudonis et~al.}{2023}]{Raudonis2023}
\begin{barticle}
\bauthor{\bsnm{Raudonis}, \binits{M.}},
\bauthor{\bsnm{Roura}, \binits{A.}},
\bauthor{\bsnm{Meister}, \binits{M.}},
\bauthor{\bsnm{Lotz}, \binits{C.}},
\bauthor{\bsnm{Overmeyer}, \binits{L.}},
\bauthor{\bsnm{Herrmann}, \binits{S.}}, \betal:
\batitle{{Microgravity facilities for cold atom experiments}}.
\bjtitle{Quantum Science and Technology}
(\byear{2023})
\doiurl{10.1088/2058-9565/ace1a3}
\end{barticle}
\endbibitem

\bibitem[\protect\citeauthoryear{Pelluet et~al.}{2024}]{Pelluet2024}
\begin{botherref}
\oauthor{\bsnm{Pelluet}, \binits{C.}},
\oauthor{\bsnm{Arguel}, \binits{R.}},
\oauthor{\bsnm{Rabault}, \binits{M.}},
\oauthor{\bsnm{Jarlaud}, \binits{V.}},
\oauthor{\bsnm{Metayer}, \binits{C.}},
\oauthor{\bsnm{Barrett}, \binits{B.}}, et al.:
{Atom interferometry in an Einstein Elevator}
(2024)
{\href{https://arxiv.org/abs/2407.07183}{{arXiv:2407.07183}}}
\end{botherref}
\endbibitem

\bibitem[\protect\citeauthoryear{Kulas et~al.}{2017}]{Kulas2017}
\begin{barticle}
\bauthor{\bsnm{Kulas}, \binits{S.}},
\bauthor{\bsnm{Vogt}, \binits{C.}},
\bauthor{\bsnm{Resch}, \binits{A.}},
\bauthor{\bsnm{Hartwig}, \binits{J.}},
\bauthor{\bsnm{Ganske}, \binits{S.}},
\bauthor{\bsnm{Matthias}, \binits{J.}}, \betal:
\batitle{{Miniaturized Lab System for Future Cold Atom Experiments in
  Microgravity}}.
\bjtitle{Microgravity Science and Technology}
\bvolume{29}(\bissue{1-2}),
\bfpage{37}--\blpage{48}
(\byear{2017})
\doiurl{10.1007/s12217-016-9524-7}
\end{barticle}
\endbibitem

\bibitem[\protect\citeauthoryear{Isichenko et~al.}{2022}]{Isichenko2022}
\begin{botherref}
\oauthor{\bsnm{Isichenko}, \binits{A.}},
\oauthor{\bsnm{Chauhan}, \binits{N.}},
\oauthor{\bsnm{Bose}, \binits{D.}},
\oauthor{\bsnm{Wang}, \binits{J.}},
\oauthor{\bsnm{Kunz}, \binits{P.D.}},
\oauthor{\bsnm{Blumenthal}, \binits{D.J.}}:
{Photonic integrated beam delivery in a rubidium 3D magneto-optical trap}.
Nature Communications
\textbf{14}(1)
(2022)
\doiurl{10.1038/s41467-023-38818-6}
{\href{https://arxiv.org/abs/2212.11417}{{2212.11417}}}
\end{botherref}
\endbibitem

\bibitem[\protect\citeauthoryear{K\"{u}rbis et~al.}{2020}]{Kurbis2020}
\begin{barticle}
\bauthor{\bsnm{K\"{u}rbis}, \binits{C.}},
\bauthor{\bsnm{Bawamia}, \binits{A.}},
\bauthor{\bsnm{Kr\"{u}ger}, \binits{M.}},
\bauthor{\bsnm{Smol}, \binits{R.}},
\bauthor{\bsnm{Peters}, \binits{A.}},
\bauthor{\bsnm{Wicht}, \binits{A.}},
\bauthor{\bsnm{Tr\"{a}nkle}, \binits{G.}}:
\batitle{Extended cavity diode laser master-oscillator-power-amplifier for
  operation of an iodine frequency reference on a sounding rocket}.
\bjtitle{Appl. Opt.}
\bvolume{59}(\bissue{2}),
\bfpage{253}--\blpage{262}
(\byear{2020})
\doiurl{10.1364/AO.379955}
\end{barticle}
\endbibitem

\bibitem[\protect\citeauthoryear{Duncker et~al.}{2014}]{Duncker2014}
\begin{barticle}
\bauthor{\bsnm{Duncker}, \binits{H.}},
\bauthor{\bsnm{Hellmig}, \binits{O.}},
\bauthor{\bsnm{Wenzlawski}, \binits{A.}},
\bauthor{\bsnm{Grote}, \binits{A.}},
\bauthor{\bsnm{Rafipoor}, \binits{A.J.}}, \betal:
\batitle{{Ultrastable, Zerodur-based optical benches for quantum gas
  experiments}}.
\bjtitle{Applied Optics}
\bvolume{53}(\bissue{20}),
\bfpage{4468}
(\byear{2014})
\doiurl{10.1364/AO.53.004468}
\end{barticle}
\endbibitem

\bibitem[\protect\citeauthoryear{Schkolnik
  et~al.}{2016}]{schkolnik_compact_2016}
\begin{botherref}
\oauthor{\bsnm{Schkolnik}, \binits{V.}},
\oauthor{\bsnm{Hellmig}, \binits{O.}},
\oauthor{\bsnm{Wenzlawski}, \binits{A.}},
\oauthor{\bsnm{Grosse}, \binits{J.}},
\oauthor{\bsnm{Kohfeldt}, \binits{A.}},
\oauthor{\bsnm{Döringshoff}, \binits{K.}}, et al.:
A compact and robust diode laser system for atom interferometry on a sounding
  rocket
\textbf{122}(8),
217
(2016)
\doiurl{10.1007/s00340-016-6490-0}
\end{botherref}
\endbibitem

\bibitem[\protect\citeauthoryear{Mihm et~al.}{}]{mihm_zerodur_2019}
\begin{botherref}
\oauthor{\bsnm{Mihm}, \binits{M.}},
\oauthor{\bsnm{Marburger}, \binits{J.P.}},
\oauthor{\bsnm{Wenzlawski}, \binits{A.}},
\oauthor{\bsnm{Hellmig}, \binits{O.}},
\oauthor{\bsnm{Anton}, \binits{O.}},
\oauthor{\bsnm{Döringshoff}, \binits{K.}}, et al.:
{ZERODUR}\textcopyright~based optical systems for quantum gas experiments in
  space
\textbf{159},
166--169
\doiurl{10.1016/j.actaastro.2019.03.060}
\end{botherref}
\endbibitem

\bibitem[\protect\citeauthoryear{Mihm}{2020}]{mihm_phd}
\begin{botherref}
\oauthor{\bsnm{Mihm}, \binits{M.}}:
Laser system technology for quantum experiments in space and beyond.
PhD thesis,
Mainz
(2020).
\doiurl{10.25358/openscience-5498}
\end{botherref}
\endbibitem

\bibitem[\protect\citeauthoryear{Toptica}{2021}]{tptc_BR-LRS-2021-04}
\begin{botherref}
\oauthor{\bsnm{Toptica}}:
Laser Rack Systems For Quantum Technology 2.0 Applications.
\url{https://www.toptica.com/fileadmin/Editors_English/11_brochures_datasheets/01_brochures/toptica_BR_Laser_Rack_Systems.pdf}
Accessed 26 September 2023
\end{botherref}
\endbibitem

\bibitem[\protect\citeauthoryear{Krischke et~al.}{2021}]{suk_fibreprotcluster}
\begin{botherref}
\oauthor{\bsnm{Krischke}, \binits{A.}},
\oauthor{\bsnm{Schulz}, \binits{M.}},
\oauthor{\bsnm{Knothe}, \binits{C.}},
\oauthor{\bsnm{Oechsner}, \binits{U.}}:
Fiber Port Cluster -- Rugged, Modular and Fiber Coupled Beam Splitting and
  Combining Units, Article{\textunderscore}Cluster 27.08.2021.
Schäfter\,+\,Kirchhoff.
\url{https://www.sukhamburg.com/documents/Article_Cluster.pdf}
Accessed 26 September 2023
\end{botherref}
\endbibitem

\bibitem[\protect\citeauthoryear{Schäfter\,+\,Kirchhoff}{2022}]{suk_shutter}
\begin{botherref}
Electro-magnetic Shutter 48EMS-6.
Schäfter\,+\,Kirchhoff.
\url{https://www.sukhamburg.com/products/details/48EMS-6}
Accessed 26 September 2023
\end{botherref}
\endbibitem

\bibitem[\protect\citeauthoryear{Thorlabs}{2019}]{thrl_oij780apc}
\begin{botherref}
\oauthor{\bsnm{Thorlabs}}:
IO-J-780APC Fiber Isolator User Guide.
\url{https://www.thorlabs.com/drawings/ec3fddd414322c7-361FED53-B586-C835-01FFCADAF113E358/IO-J-780APC-Manual.pdf}
Accessed 26 September 2023
\end{botherref}
\endbibitem

\bibitem[\protect\citeauthoryear{Gooch\,\&\,Housego}{March
  2021}]{gah_PEC0013iss8}
\begin{botherref}
Gooch\,\&\,Housego:
Material Specification and Component Reliability Response, Doc. No.
  PEC0013iss8.
(March 2021).
Gooch\,\&\,Housego. Available from the authors on request.
\end{botherref}
\endbibitem

\bibitem[\protect\citeauthoryear{Toptica}{2021a}]{tptc_prolock}
\begin{botherref}
\oauthor{\bsnm{Toptica}}:
DLC Pro Lock -- Software License for DLC Pro.
\url{https://www.toptica.com/products/tunable-diode-lasers/laser-locking-electronics/dlc-pro-lock}
Accessed 26 September 2023
\end{botherref}
\endbibitem

\bibitem[\protect\citeauthoryear{Toptica}{2021b}]{tptc_falcpro}
\begin{botherref}
\oauthor{\bsnm{Toptica}}:
FALC Pro -- Digitally Controlled Fast Laser Locking Module.
\url{https://www.toptica.com/products/tunable-diode-lasers/laser-locking-electronics/falc-pro}
Accessed 26 September 2023
\end{botherref}
\endbibitem

\bibitem[\protect\citeauthoryear{Schmidt-Eberle}{2023}]{tptc_linewidth}
\begin{botherref}
\oauthor{\bsnm{Schmidt-Eberle}, \binits{S.}}:
Application Note: Linewidth Measurement Of Diode Lasers.
Toptica.
\url{https://www.toptica.com/fileadmin/Editors_English/04_applications/10_application_notes/05_Linewidth_Measurement_of_Diode_Lasers/Linewidth-measurements-of-diode-lasers.pdf}
Accessed 16 October 2023
\end{botherref}
\endbibitem

\end{thebibliography}

\end{document}